\documentclass[letterpaper,twocolumn,10pt]{article}
\usepackage{usenix2019_v3}
\usepackage{tikz}
\usepackage{amsmath}
\usepackage{amssymb}
\usepackage{xspace}
\usepackage{breakurl}
\usepackage{url}
\usepackage{multirow}
\usepackage{pifont}
\usepackage{caption}
\usepackage{subfigure}
\usepackage{algorithm}
\usepackage[noend]{algpseudocode}
\usepackage{xcolor}
\usepackage{booktabs}
\usepackage{hyperref}
\hypersetup{
    colorlinks=true,linkcolor=black,citecolor=red
}

\newcommand{\cmark}{\ding{51}}%
\newcommand{\xmark}{\ding{55}}%
\newcommand{\paraspace}{\vspace{0.04in}}
\newcommand{\parab}[1]{\paraspace\noindent{\bf #1} }

\newcommand{\red}{\textcolor{red}}
\newcommand{\blue}{\textcolor{blue}}

\newcommand{\etc}{\textit{etc}.\xspace}
\newcommand{\ie}{\textit{i}.\textit{e}.\xspace}
\newcommand{\eg}{\textit{e}.\textit{g}.\xspace}
\newcommand{\aka}{\textit{a}.\textit{k}.\textit{a}.\xspace}

\newcommand{\sys}{Gleam\xspace}

\newcommand{\rdverbs}{\textit{verbs}\xspace}
\newcommand{\rdread}{READ\xspace}
\newcommand{\rdwrite}{WRITE\xspace}

\newcommand{\rdsend}{SEND\xspace}
\newcommand{\rdreceive}{RECEIVE\xspace}
\newcommand{\rdwqe}{WQE\xspace}
\newcommand{\lkey}{\textit{L\_key}\xspace}
\newcommand{\rkey}{\textit{R\_key}\xspace}
\newcommand{\rdcqe}{CQE\xspace}
\newcommand*\circled[1]{\tikz[baseline=(char.base)]{
            \node[shape=circle,draw,inner sep=0.1pt] (char) {#1};}}
\newcommand{\romannumerber}[1] {\romannumeral #1}
\newcommand{\envelope}{\textit{envelope}\xspace}
\newcommand{\mpibcast}{$MPI\_Bcast$\xspace}

\begin{document}

\date{}

\title{Gleam: An RDMA-accelerated Multicast Protocol for Datacenter Networks}



\author{
\rm Wenxue Li$^{1}$\thanks{Co-first authors.}~~~
\rm Junyi Zhang$^{2*}$~~~
\rm Gaoxiong Zeng$^{2}$~~~
\rm Yufei Liu$^{2}$~~~
\rm Zilong Wang$^{1}$~~~\\
\rm Chaoliang Zeng$^{1}$~~~
\rm Pengpeng Zhou$^{2}$~~~
\rm Qiaoling Wang$^{2}$~~~
\rm Kai Chen$^{1}$~~~\\
\\
$^1$Hong Kong University of Science and Technology~~~~
$^2$Huawei~~~~
}

\maketitle
\begin{abstract}
RDMA has been widely adopted for high-speed datacenter networks. However, native RDMA merely supports one-to-one reliable connection, which mismatches various applications with group communication patterns (\eg, one-to-many). While there are some multicast enhancements to address it, they all fail to simultaneously achieve optimal multicast forwarding and fully unleash the distinguished RDMA capabilities. 

In this paper, we present Gleam, an RDMA-accelerated multicast protocol that simultaneously supports optimal multicast forwarding, efficient utilization of the prominent RDMA capabilities, and compatibility with the commodity RNICs. At its core, Gleam re-purposes the existing RDMA RC logic with careful switch coordination as an efficient multicast transport. Gleam performs the one-to-many connection maintenance and many-to-one feedback aggregation, based on an extended multicast forwarding table structure, to achieve integration between standard RC logic and in-fabric multicast. We implement a fully functional Gleam prototype. With extensive testbed experiments and simulations, we demonstrate Gleam's significant improvement in accelerating multicast communication of realistic applications. For instance, Gleam achieves 2.9$\times$ lower communication time of an HPC benchmark application and 2.7$\times$ higher data replication throughput.
\end{abstract}

\section{Introduction} \label{intro}

Datacenter applications impose increasingly stringent requirements for network communication, such as persistently high throughput, ultra-low latency ($\mu s$ scale), and low CPU overhead (to cut OpEx). To meet it, Remote Direct Memory Access (RDMA) is emerging as the de-facto networking technology going beyond 40Gbps links. Many tech giants have adopted RDMA into their production datacenters~\cite{gao2021cloud, zhu2015congestion, mittal2015timely, guo2016rdma}. These datacenters host various network-intensive applications, such as deep learning~\cite{li2014communication, jiang2020unified}, cloud storage~\cite{gao2021cloud, kalia2014using}, graph exploration~\cite{shi2016fast}, \etc, which benefit greatly from the underlying RDMA communication.

However, native RDMA transports only support one-to-one reliable connection (RC)~\cite{rocev2}, which mismatches various applications with group communication patterns~\cite{gao2021cloud, dean2004mapreduce, li2014communication} (\eg, one-to-many). Multicast is prevalent in datacenter applications ($\S$\ref{multicast-pattern}). Specifically, previous work shows that multicast is the top-2 communication pattern in a High-performance Computing (HPC) cluster~\cite{mpiusage}. In High-performance Linpack (HPL)~\cite{hpl}, an HPC's benchmark application, more than 90\% communication traffic is the multicast pattern.  

To meet applications' requirements, a straightforward solution is to provide a multicast primitive. However, while there are some existing multicast solutions~\cite{openmpi, nccl, shahbaz2019elmo, Infiniband} ($\S$\ref{multicast}), none of them simultaneously achieve two of the performance requirements: 1)~optimal forwarding multicast traffic; 2)~fully unleashing the distinguished RDMA capabilities.

On one hand, some primary distributed frameworks (\eg, MPI~\cite{openmpi} and NCCL~\cite{nccl}) choose to develop their private multicast protocols upon the RDMA one-to-one RC transport (\aka, application-layer multicast). Thus they can efficiently utilize the prominent RDMA capabilities. However, application-layer multicast cannot achieve optimal traffic forwarding, resulting in inefficient bandwidth utilization and communication bottleneck~\cite{diab2022orca, shahbaz2019elmo}. Although many multicast algorithms are proposed to mitigate the bottleneck (\eg, Ring and Double-binary tree~\cite{gibiansky2017bringing, mlsl, oneccl}), they inevitably incur longer communication distance and higher latency resulting from intermediate nodes' data forwarding. 

On the other hand, in-fabric multicast can achieve optimal multicast forwarding with efficient bandwidth utilization and minimized communication distance. However, the in-fabric multicast cannot benefit from advanced RDMA capabilities. In particular, IP-based multicast~\cite{crowcroft1988multicast, diab2022orca, shahbaz2019elmo} only supports layer-3 routing without any transport-layer functionality, thus is incompatible with the RDMA RC transport. IB multicast, defined in IB standard~\cite{Infiniband}, only supports UD transport. Therefore, applications that adopt IB multicast cannot utilize the advanced one-sided \rdwrite\footnote{The RDMA \rdread doesn't fit into the multicast context.}, the extended message size, and the hardware-supported reliability. 

In addition to the performance requirements, the multicast solution should also meet the practical deployment requirements in production environments. The fundamental challenge comes from the commodity RDMA Network Interface Cards (RNICs). Commodity RNICs provide limited capability for flexible programming~\cite{gao2021cloud, zhu2015congestion}. For instance, the complete network stacks of \cite{cx5} are statically built-in with specialized circuits.

In this work, we focus on simultaneously (\romannumerber{1})~supporting the optimal multicast forwarding, (\romannumerber{2})~fully exploiting the prominent features of RDMA, and (\romannumerber{3})~satisfying the deployment requirements. To this end, we propose \sys, an RDMA-accelerated multicast solution for datacenter networks ($\S$\ref{design}). Firstly, \sys inherits the classical in-fabric distribution manner~\cite{crowcroft1988multicast, diab2022orca, shahbaz2019elmo} to achieve optimal multicast forwarding. Secondly, \sys re-purposes the existing RDMA RC logic with careful switch coordination to process multicast traffic, unleashing RDMA's superior competencies. Finally, as Gleam reuses the standard RC transport, it is compatible with the large-scale deployed commodity RNICs.

The key challenge behind Gleam is how to integrate the optimal-forwarded multicast traffic with the existing RC logic ($\S$\ref{key-chalge}). Specifically, there are two main issues. Firstly, the current connection-oriented logic~\cite{rocev2} is targeted for one-to-one connection, which cannot directly support one-to-many data delivery. Secondly, the existing reliability logic~\cite{zhu2015congestion,guo2016rdma} is designed for single feedback (including ACK, NACK, CNP, \etc) stream from a single receiver; thus, multiple feedback streams in multicast can confuse it and degrade the overall performance.  

To reuse the connection-oriented logic of RC ($\S$\ref{connection-handle}), \sys elaborately extends the widely-adopted multicast forwarding table structure~\cite{crowcroft1988multicast}. Then, based on the extended table, \sys replaces the connection-related header fields in multicast data packets to match different QPs/connections\footnote{We use QP and connection equivalently in this paper.} on different receivers. As a result, \sys achieves a virtual one-to-many connection among multicast members, providing applications with the advanced capability of RDMA connected transport service.


To reuse the reliability logic of RC ($\S$\ref{ack-aggregation}), \sys performs the many-to-one feedback aggregation in the fabric. In particular, \sys aggregates ACK so that the sender receives a unicast-like ACK stream, which is compatible with the existing ACK-interpretation logic. Moreover, \sys carefully filters NACK packets to enable the sender to correctly detect and retransmit the lost packet. Consequently, \sys can reuse the standard reliability logic and provides hardware-based reliability.

We implement a fully functional \sys switch ($\S$\ref{imple}), connected to four commodity servers. Each server is equipped with an unmodified commodity RNIC. \sys is evaluated through extensive testbed experiments and simulations ($\S$\ref{eva}). The testbed experiments show that \sys accelerates the multicast pattern by up to 2.2$\times$, and improves the realistic application's performance, such as 2.9$\times$ lower HPL communication time and 2.7$\times$ higher data replication throughput. Meanwhile, large-scale simulations demonstrate that \sys can maintain a high performance in large-scale topologies and a satisfying goodput when packet loss occurs.

\begin{figure}[t]
    \centering
    \includegraphics[width=\linewidth]{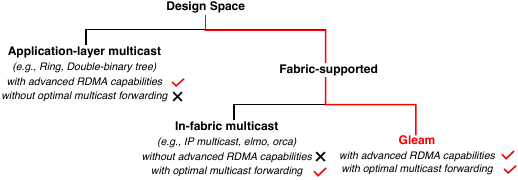}
    \caption{Design space for multicast solutions.}
    \label{fig:designspace}
    \vspace{-0.5cm}
\end{figure}

We summarize the design space for multicast solutions in Fig. \ref{fig:designspace}. \textit{To the best of our knowledge, this is the ﬁrst work to completely leverage RDMA capabilities to empower multicast communication.} The key contributions of this work are summarized below:
\begin{itemize}
\vspace{-0.2cm}
\item We observe that the reliable multicast delivery can be efficiently realized through fitting the multicast traffic into RDMA's performant RC transport.\vspace{-0.2cm}
\item We design \sys{}, an RDMA-accelerated multicast protocol that simultaneously supports optimal multicast forwarding, advanced capabilities of RDMA, and compatibility with commodity RNICs. \sys{} abstracts a virtual one-to-many connection and aggregates feedback in the middle fabric, which addresses the incompatibility between multicast traffic and existing RC logic.\vspace{-0.2cm}
\item We implement a fully functional \sys prototype. Extensive experiments demonstrate \sys's significant acceleration in multicast communication.\vspace{-0.2cm}
\end{itemize}

\section{Background and Motivation}
We first introduce some basic concepts of RDMA ($\S$\ref{rdma}); then present the prevalence of multicast pattern ($\S$\ref{multicast-pattern}); finally reveal the insufficiencies of existing solutions ($\S$\ref{multicast}). 

\subsection{RDMA Overview} \label{rdma}
RDMA is an emerging hardware-offloaded transport that implements the transport functionalities entirely in RNICs, including the basic packet encap/decap, reliability enhancements, congestion control, \etc As a consequence, RDMA provides high throughput, low latency, and efficient CPU utilization~\cite{zhu2015congestion}.

\parab{RDMA Transport Services.}
Commodity RNICs support three transport services: \textit{reliable connection} (RC), \textit{unreliable connection} (UC), and \textit{unrelibale datagram} (UD), each of which supports a different subset of RDMA operations. There are two types of operations: one-sided (\rdwrite and \rdread) and two-sided (\rdsend/\rdreceive). The one-sided operations don't involve the remote CPU, while the two-sided operations require the remote CPU to participate. 

Among the supported RDMA transport services, RC owns various advantages over UC and UD~\cite{dragojevic2014farm,monga2021birds}. Firstly, RC supports all RDMA operations, especially the one-sided \rdwrite, while UD only supports \rdsend/\rdreceive. Additionally, RC provides hardware-supported reliability, which reduces the software overhead, while UC and UD lack this benefit. Finally, the message size limits are also different; the message of RC and UC can span multiple packets with a total size of up to 2GB, while the message of UD is limited to a single packet. We summarize the RDMA transport service characteristics in Table~\ref{tab:rdma-transport}.
 

\parab{Queue Pair and Memory Region.}
RDMA applications communicate through \textit{queue pairs} (QPs) and request a network communication by submitting a \textit{work queue element} (\rdwqe) to the associated QP via \rdverbs. Each QP is identified by a QP number (QPN) and includes two queues: a \textit{send queue} (SQ) and a \textit{receive queue} (RQ). Each QP is usually associated with a \textit{completion queue} (CQ), which is used for RNIC to post the \textit{completion queue event} (\rdcqe) that contains the completion status of the previous submitted \rdwqe. 

RDMA applications register the to-be-accessed memory area as \textit{memory region} (MR) before communication starts. Each MR is associated with a \textit{virtual address} (VA) and a pair of keys: \lkey and \rkey. The application utilizes the \lkey (\rkey) as authorization to access the local (remote) MR. As the one-sided operations don't involve the remote CPU, the related MR info is formatted into the header of the one-sided request's first packet.

\subsection{Multicast Pattern}\label{multicast-pattern}
Datacenter applications are usually deployed in a distributed approach to gain more computation capability. The participants frequently communicate with others, representing group communication patterns. Multicast is a prevalent communication pattern that many applications exhibit, such as database~\cite{gao2021cloud, miao2022luna}, telemetry and monitoring system~\cite{massie2004ganglia}, HPC~\cite{top500} and machine learning system~\cite{li2014communication, jiang2020unified}. We present examples in HPC and storage networks in detail below.

\parab{HPC.} High-performance Linpack (HPL)~\cite{hpl} is a benchmark application used to rank the supercomputer's computing capacity~\cite{top500}. HPL solves a linear system, $Ax=b$ (usually with a large order), by LU decomposition. The overall HPL procedure is divided into multiple epochs, each including three steps:  \textit{Panel Factorization}, \textit{Panel Broadcast} and \textit{Update}. The \textit{Panel Broadcast} is a standard multicast transmission, and the \textit{Update} includes a communication phase (called \textit{Row Swap}) which can be implemented with multicast. As the volume of communicated traffic is large~\cite{dongarra2003linpack}, an efficient multicast can significantly improve HPL's performance. 

\parab{Storage network.} Distributed storage networks offer high availability and durability using multiple replications~\cite{gao2021cloud}. Failures of storage devices are inevitable in large-scale clusters. Therefore, replicas are critical to prevent data loss.
The replication delivery impacts application's QoS significantly, as it may last for minutes~\cite{miao2022luna}. Thus, an efficient multicast service can substantially improve the application's experience.


\begin{figure}[t]
	\centering
	\subfigure[Multiple unicasts.]
		{\includegraphics[width=0.3\columnwidth]{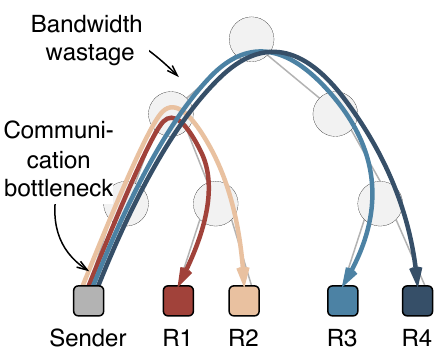}
		\label{fig:intro:unicast}}
	\subfigure[Overlay multicast.]
		{\includegraphics[width=0.3\columnwidth]{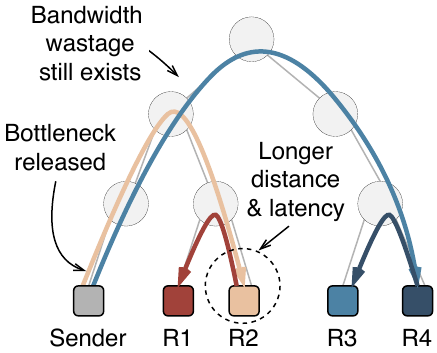}
		\label{fig:intro:overlay}}
	\subfigure[In-fabric multicast.]
		{\includegraphics[width=0.3\columnwidth]{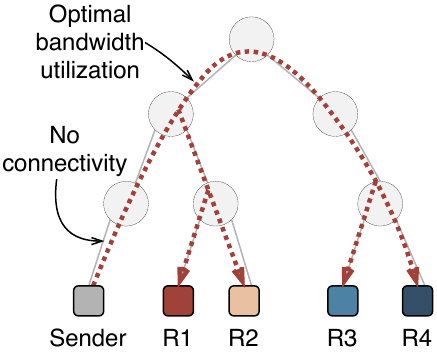}
		\label{fig:intro:infabric}}
	\vspace{-0.25cm}
	\caption{Existing multicast solutions.}
\end{figure}

\begin{table}[t]
	\small
    \centering
    \begin{tabular}{c|c|c|c|c}
    \toprule
		\textbf{} & \textbf{SEND/RECV} & \textbf{WRITE} & \textbf{READ}  & \textbf{Message Size}\\
    \midrule
    	\textbf{RC}	& \blue{\cmark}	& \blue{\cmark}	 & \blue{\cmark} & 2\,GB \\ 
    	\textbf{UC}	& \blue{\cmark}	& \blue{\cmark}	 & \red{\xmark} & 2\,GB \\
		\textbf{UD}	& \blue{\cmark}	& \red{\xmark}	 & \red{\xmark} & 4\,KB \\
    \bottomrule
    \end{tabular}
    \caption{RDMA operations and message size that supported in each RDMA transport.}
    \label{tab:rdma-transport}
    \vspace{-0.25cm}
\end{table}

\subsection{Insufficiencies of Existing Solutions} \label{multicast}
Existing multicast solutions cannot simultaneously achieve optimal multicast forwarding and efficient utilization of advanced RDMA capabilities.

\parab{Application-layer Multicast.} 
Primary distributed computation frameworks, such as MPI~\cite{openmpi} and NCCL~\cite{nccl}, develop their private multicast protocol upon the RC transport service. Thus they can efficiently utilize the advanced features of RC. However, they cannot achieve optimal traffic forwarding resulting in inefficient communication performance. There are primarily two types of implementations: 1) the basic multiple sender-to-receiver unicasts (abbreviated as multiple unicasts in this paper), and 2) the overlay multicast. 

As illustrated in Fig.~\ref{fig:intro:unicast}, for multiple unicasts, the sender establishes multiple connections and transmits identical data to different receivers via multiple unicasts. This inevitably incurs severe bandwidth wastage and leads to a potential network bottleneck at the sender side. To address it, various overlay algorithms, \eg, Ring and Double-binary tree~\cite{gibiansky2017bringing, mlsl, oneccl}), are proposed to leverage a pipelined transmission strategy. Specifically, some intermediate receivers can relay data distribution after receiving data, thus spreading the transmission burden to all participants (Fig.~\ref{fig:intro:overlay}). However, the overlay multicast results in longer communication distance and higher latency because of intermediate nodes' forwarding. Specifically, in intermediate nodes, packets must go through \textsf{RX} stack, interrupt the CPU to make a forwarding decision, and finally go through \textsf{TX} stack to be sent out.

\parab{In-fabric Multicast.}
With in-fabric multicast, the sender only needs to send out one single copy of data; then, the fabric replicates and forwards the data to multiple receivers via a multicast distribution tree (Fig.~\ref{fig:intro:infabric}). However, existing solutions cannot unleash the full power of RDMA. In particular, IP-based multicast~\cite{crowcroft1988multicast, diab2022orca, shahbaz2019elmo} only supports layer-3 routing without any layer-4 functionality. Developing a specific layer-4 protocol in software upon existing IP-based multicast incurs additional CPU overhead and still forgo the efficient RDMA. IB multicast, defined in IB standard~\cite{Infiniband}, only supports UD transport, inheriting all constraints of UD transport, as described in $\S$\ref{rdma}. Therefore, the insufficiencies of existing in-fabric multicast solutions make their large-scale deployment less attractive.

\section{\sys}\label{design}

We first describe the key ideas and design challenges of \sys ($\S$\ref{key-chalge}). Then we overview the \sys structure ($\S$\ref{design-overview}), briefing its main components and how they work together. After that, we elaborate these design components one by one ($\S$\ref{connection-handle}-$\S$\ref{other-design}). 

\subsection{Key Ideas and Challenges} \label{key-chalge}

\sys focuses on simultaneously achieving (\romannumerber{1})~the optimal multicast forwarding, (\romannumerber{2})~the efficient utilization of the advanced capabilities of RDMA, and (\romannumerber{3})~satisfying the deployment requirements. To this end, the key ideas behind \sys are to (\romannumerber{1})~perform the optimal multicast forwarding in the in-fabric distribution manner~\cite{crowcroft1988multicast, diab2022orca, shahbaz2019elmo}; and (\romannumerber{2})~re-purpose the native RDMA RC logic with careful switch coordination for an efficient multicast transport. 

However, there are critical challenges blocking the way: \textit{how to achieve integration between the optimal multicast forwarding and the existing RDMA RC logic}. Specifically, there are two main compatibility issues.

\begin{figure}[t]
	\centering
	\subfigure[Discarding packets.]
		{\includegraphics[width=0.45\columnwidth]{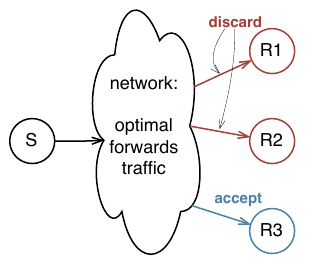}
		\label{fig:design:incompt-net}}
	\subfigure[Inside $R_1$'s RNIC.]
		{\includegraphics[width=0.45\columnwidth]{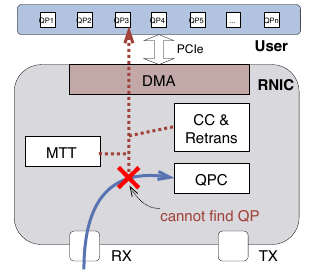}
		\label{fig:design:incompt-detail}}
	\vspace{-0.25cm}
	\caption{Traditional-forwarded traffic is incompatible with the existing connection-oriented logic of RDMA. $R_1$ and $R_2$ cannot find the associated QPs, so they discard packets. The dotted lines in (b) mean non-reachable.}
	\vspace{-0.25cm}
	\label{fig:design:incompt}
\end{figure}

The first is that the connection-oriented logic~\cite{rocev2} of RC cannot find the associated QPs when receiving the traditional-forwarded multicast packets, illustrated in Fig.~\ref{fig:design:incompt}. During the traditional multicast forwarding, switches won't change the packet's layer-4 header. Switches either copy the entire packet~\cite{crowcroft1988multicast, Infiniband} or only modify the layer-3 (IP) header for layer-3 multicast routing~\cite{diab2022orca, shahbaz2019elmo}. Thus all packets contain an identical layer-4 header, which only matches at most one connection. The non-matched RNIC will discard these packets as it cannot find the associated QP and Queue Pair Context (QPC) based on the non-matched layer-4 header. 

Secondly, even if receivers can accept the packets and find the associated QPs, the second incompatibility impeding the leverage of RC is the existing reliability logic~\cite{zhu2015congestion, guo2016rdma}. The standard reliability logic is designed for single feedback (including ACK, NACK, CNP, \etc) stream from a single receiver. Thus, multiple feedback streams from multiple receivers can confuse RC and disturb its loss detection and retransmission routines. 

\sys is a fabric-supported multicast protocol that substantially differs from the traditional layer-3 in-fabric approaches. \sys systematically integrates its design components to address these two incompatibilities.


\begin{figure*}[t]
    \centering
    \includegraphics[width=0.95\linewidth]{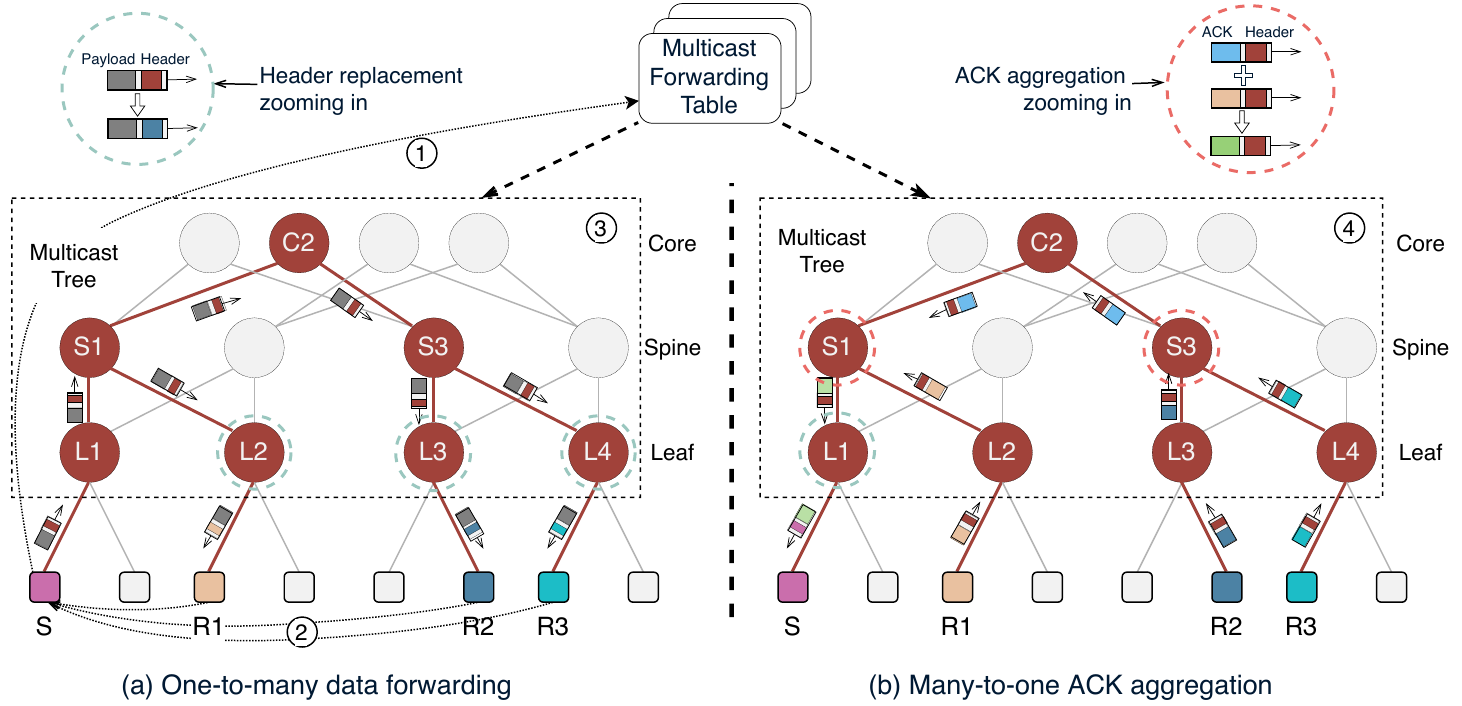}
    \caption{\sys Architecture. We take a three-layer fat-tree topology as an example. The sender $S$, the three receivers, $R_1$, $R_2$, and $R_3$, and the colored switches form a multicast group. The left and right sub-figures illustrate the one-to-many data forwarding and the many-to-one feedback aggregation (ACK as the example in the figure), respectively.}
    \label{fig:overview}
    \vspace{-0.5cm}
\end{figure*}

\subsection{\sys Overview} \label{design-overview}
Fig.~\ref{fig:overview} illustrates the architecture of \sys. The working steps of \sys are composed of three main phases: the control-plane multicast group registration (\circled{1} $\&$ \circled{2}), the data-plane one-to-many data forwarding (\circled{3}), and the data-plane many-to-one feedback aggregation (\circled{4}).

For the control-plane multicast group registration, \sys elaborately extends the legacy multicast forwarding table structure by integrating layer-4 states. We develop an out-of-band UDP-based protocol called \envelope to register forwarding table to switches—the master node\footnote{For ease of understanding, readers can think master node as the multicast source. Actually, the master node can be any node in the multicast group.} in the multicast group collects the layer-4 states of other nodes, fits these states into the \envelope packet format, and transmits to switches and other nodes for building forwarding table and affirming the multicast membership, respectively (\circled{1}). The involved nodes will answer ACKs to the master node to confirm its participation (\circled{2}). Due to space limitations, we present the extended forwarding table structure in the main body, leaving the remaining control-plane details in Appendix~\ref{apx:regis}.

For the data-plane one-to-many data forwarding ($\S$\ref{connection-handle}), \sys reserves the optimal multicast forwarding and achieves a virtual one-to-many connection. Let's take the multicast communication in Fig.~\ref{fig:overview} for an example. $S$ only transmits data once via the existing RC connection. Then the switches in the multicast tree copy data and forward them to multiple receivers via the optimal paths. In addition, some specific switches replace the connection-related states in the packet header to match different QPs in different receivers (\circled{3}), thus achieving a virtual one-to-many connection. For instance, $L_1$, $S_1$, $C_2$, and $S_3$ copy and forward data packets to specific ports that are identified in the forwarding table. Additionally, $L_2$, $L_3$, and $L_4$ replace some connection-related states in packet header to match QPs in $R_1$, $R_2$, and $R_3$.

After receiving data packets, receivers, $R_1$, $R_2$, and $R_3$, generate normal ACK/NACK/CNP packets following the existing RC logic. Then the fabric aggregates ACK, filters NACK/CNP, and forwards these feedbacks to the sender (\circled{4}) ($\S$\ref{ack-aggregation} $\&$ $\S$\ref{other-design}). We take ACK as an example. As shown in Fig.~\ref{fig:overview}, $L_2$, $L_3$, $L_4$, and $C_2$ only forward ACK packets to next-hop switches, as there is only one ACK stream as input. $S_3$ and $S_1$ perform ACK aggregation, as there are multiple ACK streams as input. $L_1$ changes the connection-related states in the ACK header to match the $S$'s QP before forwarding the aggregated ACK. 

The aggregated ACK and NACK packets enable the sender to transmit the subsequent new data packets or retransmit the lost ones. The filtered CNPs are used to adjust the sender's sending rate. Every data packet will go through the above four steps until the multicast communication job is finished.

\subsection{One-to-many Data Forwarding} \label{connection-handle}

For the data-plane one-to-many data forwarding, \sys firstly reserves the optimal multicast forwarding, adopted by the previous in-fabric multicast solutions~\cite{crowcroft1988multicast, diab2022orca, shahbaz2019elmo}. Thus the sender only needs to send one copy of data, and the fabric makes multiple copies and forwards them to multiple receivers via the optimal paths. Therefore, there is no bandwidth wastage, and the communication distance and latency are minimized. 

\begin{figure}[t]
    \centering
    \includegraphics[width=0.95\linewidth]{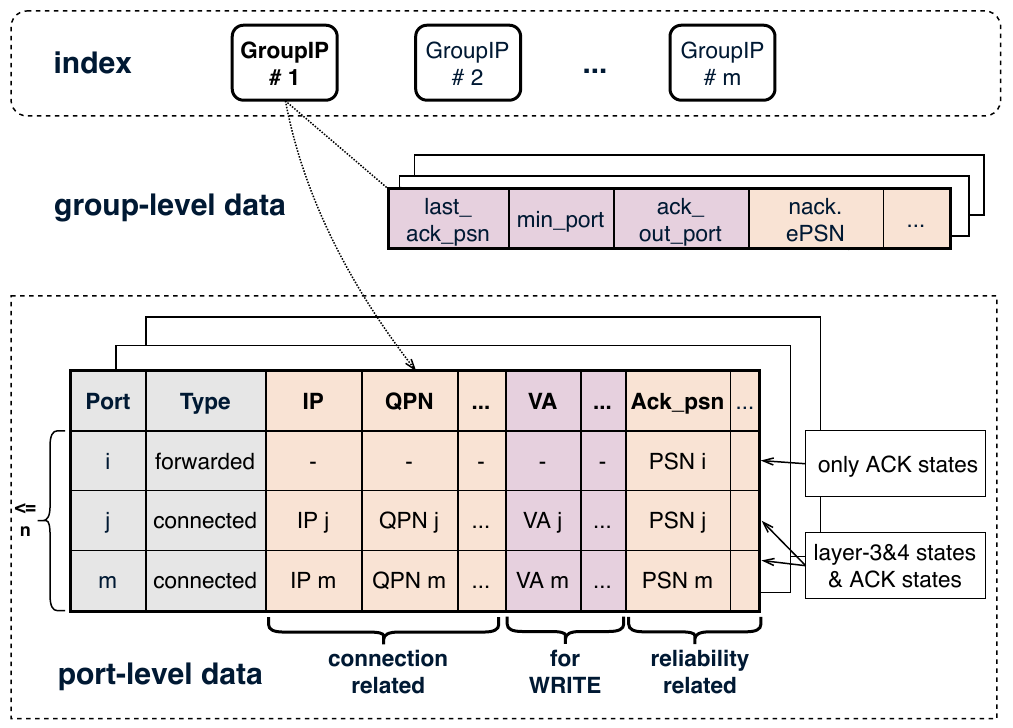}
    \vspace{-0.25cm}
    \caption{The extended multicast forwarding table in \sys.}
    \label{fig:table}
    \vspace{-0.5cm}
\end{figure}

Secondly, departing from the previous works that only support layer-3 routing, \sys integrates layer-4 states into the fabric and further achieves a virtual one-to-many connection. Because of this, the single connected QP in the sender can simultaneously communicate with multiple QPs on multiple receivers. Therefore, \sys can fully leverage the advanced features of connected transport service, \ie, the one-sided \rdwrite operation and extended message size, as discussed in $\S$\ref{rdma}. Besides the more efficient bandwidth utilization, \sys outperforms the application-layer multicast with shorter communication distance and lower forwarding latency, as there is no additional \rdwqe and \rdcqe processing, and the intermediate nodes are not involved in forwarding data.

\parab{Extended multicast forwarding table.} \sys elaborately extends the traditional multicast forwarding table structure by integrating layer-4 states. The extended table, illustrated in Fig.~\ref{fig:table}, is the foundation of \sys. The indexed key of the table is the \textit{multicast group IP address} (abbreviated as GroupIP). Many multicast groups can exist simultaneously, each with a unique GroupIP.

\begin{figure}[t]
    \centering
    \includegraphics[width=\linewidth]{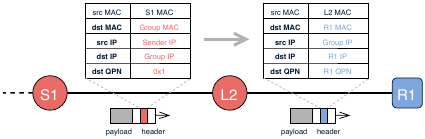}
	\vspace{-0.5cm}
    \caption{The packet header change in $S_1$-$L_2$-$R_1$ path in Fig. \ref{fig:overview}. The $dest\_IP$ and $dest\_QPN$ are changed from GroupIP and $0x1$ to $R_1$'s IP and QPN. \sys changes the source IP from the sender's IP to GroupIP and manually replaces the destination MAC address.}
    \label{fig:header-change}
    \vspace{-0.5cm}
\end{figure}

The GroupIP indexes two types of states: the group-level states and the port-level states. The group-level states contain statistics of the multicast group, which are used for the many-to-one feedback aggregation. The port-level states are formatted into an array with at most $n$ ( \# of switch ports) entries, only containing ports included in the multicast tree. The entry with $port$ as $i$ represents states related to $port_i$. Each entry is assigned one of two types: $connected$ and $forwarded$, where $connected$ means that this port is directly connected to a receiver node, and $forwarded$ means that the next hop is a switch. The $connected$ entry contains the connected receiver's layer-3 and layer-4 states, as well as the ACK/NACK states, and the $forwarded$ entry only contains ACK/NACK states. $\S$\ref{ack-aggregation} introduces the usage of these ACK/NACK states. 

The specific memory space used by one multicast table depends on the number of ports involved in the multicast group, which is at most $n$. We calculate that 1K multicast groups at most cost 0.92MB memory when each group contains the maximum entries. Moreover, the design goal of \sys is not to compress the switch-maintained states but to provide a general multicast protocol with prominent RDMA features. We can use many approaches to extend \sys to support more groups, and we'll talk about them in $\S$\ref{works}.

\parab{Establishing QPs.} Each multicast member follows the common unicast-like steps to establish the RC QP but assigns a virtual destination to it. Specifically, the destination IP is set as a unique GroupIP, and the destination QPN can be assigned as any non-conflicting value (\eg, $0x1$). Commodity RNICs provide the application with the programming interface to specify the destination IP and QPN without modifying the RNIC circuit~\cite{qpmodi}. After QPs establishment, multicast members exchange their QPs information and register the above-described forwarding table to switches, as described in Appendix~\ref{apx:regis}. Once the registration finishes, the sender can start sending multicast data packets.


\parab{One-to-many data forwarding.} Switches involved in the multicast tree are responsible for forwarding data via the multicast tree and modifying packet headers to match different QPs. The switch follows Algorithm~\ref{alg:data-forward} to process data packets. Upon receiving a data packet ($p$), the switch uses the destination IP ($p.dest\_IP$) in the packet header to index the associated multicast forwarding table ($T$). Then switch iterates all entries (one entry corresponds to one port) in $T$ and actions as follows: (\romannumerber{1}) if the type is $forwarded$: creates a packet copy and forwards it through this port; (\romannumerber{2}) if the type is $connected$: creates a packet copy, modifies its connection-related states, and forwards it through this port.

We take one path, $S_1$ to $L_2$ to $R_1$, of the multicast tree in Fig.~\ref{fig:overview} as an example to show the header change, which is illustrated in Fig.~\ref{fig:header-change}. Firstly, the destination IP and QPN are modified to match $R_1$'s QP identification, as described in $\S$\ref{key-chalge}. Besides, \sys changes the source IP from the sender's IP to GroupIP. As a result, when $R_1$ generates feedback, the feedback's destination IP will be the data packet's source IP, \ie, GroupIP. Thus feedback packets can also index the associated forwarding table by their destination IP. Besides, \sys switches replace the destination MAC address to avoid the receiver's MAC layer discarding the packet. 


\begin{algorithm}[t]
\caption{Forwarding Packets and Replacing Headers.}\label{alg:data-forward}
\begin{algorithmic}[1]
\State $p\gets $ data packet
\State $j\gets$ port that $p$ enters
\State $T\gets $ multicast forwarding table indexed by $p.dest\_IP$
\For{$Entry \in T$}\Comment{\textcolor{gray}{loop over $T$ and copy/forward $p$}}
	\If{$Entry.type = forwarded$ \& $Entry.port \neq j$}
		\State $\overline{p}\gets$ a copy of $p$
		\State send $\overline{p}$ out from $Entry.port$
	\EndIf
	\If{$Entry.type = connected$ \& $Entry.port \neq j$}
		\State $\overline{p}\gets$ a copy of $p$
		\State $\overline{p}.dest\_IP(QPN) =Entry.dest\_IP(QPN)$
		\State $\overline{p}.(...)=Entry.(...)$
		\State send $\overline{p}$ out from $Entry.port$
	\EndIf
\EndFor
\end{algorithmic}
\end{algorithm}

\parab{Support for one-to-many \rdwrite.}
\sys maintains connectivity between the sender and multiple receivers, which is sufficient for \rdsend/\rdreceive. However, \rdwrite requires more support. \rdwrite allows a node to write a memory slot on a remote node. The to-be-written MR info (including the remote VA and \rkey) is indicated in the first packet of the \rdwrite request. The \rdwrite responder's RNIC will check the MR info and execute the request only when they are correct. Otherwise, the packets will be discarded. To enable one-to-many \rdwrite, \sys needs to modify the MR states in the \rdwrite request header for different receivers. 

Besides maintaining MR info for different receivers, the MR info needs to be updated for every \rdwrite request because the MR changes with different \rdwrite requests. We force the host application to invoke an extra \rdwrite message which contains the MR states of different receivers, before submitting the actual \rdwrite request. Then the leaf switch recognizes this special message, updates MR info to the table, and replaces the MR states for the subsequent real \rdwrite request. This per-request updating scheme introduces minimal extra bandwidth overhead as long as the extra \rdwrite message is small compared to the total volume of transmitted data. We evaluate the performance of one-to-many \rdwrite in $\S$\ref{eval:storage}. Moreover, we discuss a possible way to avoid this extra overhead in Appendix~\ref{apx:opwrite}. This alternative approach requires the RNIC's modification.

\subsection{Many-to-one Feedback Aggregation} \label{ack-aggregation}

\begin{algorithm}[t]
\caption{Many-to-one ACK/NACK Aggregation}\label{alg:ack-aggre}
\begin{algorithmic}[1]
\State $p\gets $ ACK/NACK packet
\State $T\gets $ forwarding table indexed by $p.dest\_IP$
\State $Entry\gets$ entry in $T$ with $port$ $i$ that $p$ enters
\State $min\_port\gets$ port with minimum $ack\_psn$ last time
\State $last\_ack\_psn\gets$ last aggregated ACK's PSN
\If{$p$ is $ACK$}
\If{$p.psn \geq Entry.ack\_psn$}\Comment{\textcolor{gray}{update $Entry$}}
		\State $Entry.ack\_psn = p.psn$
		\State $Entry.(...) = p.(...)$
	\EndIf
	\If{$i = min\_port$ \& $p.psn \geq last\_ack\_psn$}\label{line:trigger}
		\State \Call{GenerateNewAck/Nack}{ }
	\EndIf
\Else \Comment{\textcolor{gray}{$p$ is a NACK packet}}
	\If{$p.psn - 1 \geq Entry.ack\_psn$}
		\State $Entry.ack\_psn = p.psn - 1$\Comment{\textcolor{gray}{update $Entry$}}
	\EndIf
	\If{$p.psn \leq T.nack.ePSN$} \label{line:nackupdate}\Comment{\textcolor{gray}{update $T.nack$}}
		\State $T.nack.ePSN = p.psn$
		\State $T.nack.(...) = p.(...)$
		\State \Call{GenerateNewAck/Nack}{ }
	\EndIf
\EndIf
\end{algorithmic}
\end{algorithm}

With the extended one-to-many data forwarding, \sys can optimally forward data replicas to multiple receivers, and different receivers' QPs can accept the packet. However, the second incompatibility impeding the leverage of RC is the existing reliability logic~\cite{zhu2015congestion, guo2016rdma}. The current reliability logic in commodity RNICs is designed to interpret a single feedback stream from a single receiver; thus, multiple feedback streams from multiple receivers can compromise reliability. 

Feedback contains various types of packets, such as ACK, NACK, notification packets for congestion control (CC) (such as CNP~\cite{zhu2015congestion}), \etc We talk about ACK and NACK here, and the processing for CC-related feedback is described in $\S$\ref{other-design}. \sys performs the in-fabric many-to-one ACK aggregation and NACK filtering to deliver a unicast-like ACK/NACK stream to the sender. As a result, the sender can correctly interpret ACK and proceeds with data transmission. Besides, when the loss occurs, the sender can precisely detect and retransmit the lost packet.

The basic principles of ACK-aggregation/NACK-filtering are that (\romannumerber{1}) the multicast source should receive an ACK only when \textit{\textbf{all}} receivers have received the corresponding packets; (\romannumerber{2}) the multicast source should receive a NACK when \textit{\textbf{any}} receiver loses a packet. Moreover, the aggregation needs to consider the processing rules of RDMA protocol, such as the go-back-N retransmission and ACK coalescing. 

\begin{algorithm}[t]
\caption{ACK/NACK Generation}\label{alg:generation}
\begin{algorithmic}[1]
\Function{GenerateNewAck/Nack}{ }
\State $T\gets $ forwarding table indexed by $p.dest\_IP$
\State $ack\_out\_port\gets$ port that data packets enter
\State $min\_psn\gets \infty$
\State $min\_port\gets -1$
\For{$Entry \in T$} \Comment{\textcolor{gray}{find the  minimized $ack\_psn$}}
		\If{$min\_psn < Entry.ack\_psn$}
			\State $min\_psn = Entry.ack\_psn$
			\State $min\_port = i$
		\EndIf
\EndFor
\State \textcolor{gray}{// generate aggregated ACK}
\State create ACK packet $p$ with ($psn = min\_psn$)
\State send $p$ through $ack\_out\_port$ 
\State \textcolor{gray}{// check for generating NACK}
\If{$min\_psn\ + 1 = T.nack.ePSN$} \label{line:nack}
	\State create NACK packet $p$ 
	\State $p.psn$ = $T.nack.ePSN$
	\State send $p$ through $ack\_out\_port$
\EndIf
\State update global $last\_ack\_psn$ with $min\_psn$
\EndFunction
\end{algorithmic}
\end{algorithm}

\sys maintains ACK/NACK-related information in the extended multicast forwarding table, illustrated in Fig.~\ref{fig:table}. Upon receiving an ACK/NACK packet, switches find the associated forwarding table by ACK/NACK packets' destination IP, \ie, GroupIP. We first present a working logic without packet loss and then describe how to handle NACK packets. 

\parab{Handle ACK.} The ACK-related states includes (\romannumerber{1})~the group-level data, including the PSN of the last aggregated ACK ($last\_ack\_psn$), the port from which the ACK should be sent ($ack\_out\_port$), \etc; (\romannumerber{2})~the port-level data, including the largest acked PSN ($ack\_psn$) of each port. Switches process ACK packets, update the related states in the forwarding table, and generate the aggregated ACK packet, following Algorithm~\ref{alg:ack-aggre}. 

Upon receiving an ACK packet, the switch firstly updates this port's $ack\_psn$ if the PSN of incoming ACK is larger than the old $ack\_psn$. Then, if the trigger condition (Line~\ref{line:trigger}) is satisfied, the \textsc{GenerateNewAck/Nack} function in Algorithm~\ref{alg:generation} is called to generate an aggregated ACK. The aggregated ACK contains the minimum $ack\_psn$ recorded by the switch, which is found by iterating all multicast forwarding table entries. As a result, each aggregated ACK forwarded by the switch confirms that all downstream receivers have received the corresponding data packets. 

The port that owns the minimum $ack\_psn$ is recorded as $min\_port$. Each time an ACK with a larger PSN is received from $min\_port$ (Line~\ref{line:trigger}), the ACK generation is triggered. Thus not every ACK packet will trigger the generation, and the number of ACKs received by the source is reduced.

\parab{Handle NACK.} When the loss occurs, the receiver generates NACK packet to notify the multicast sender. The NACK packet ($p$) contains the receiver's expected PSN ($p.psn$). Each NACK will acknowledge all data packets with PSN smaller than the expected PSN. This rule must be carefully handled in NACK generation. 

\begin{figure}[t]
	\centering
	\includegraphics[width=\linewidth]{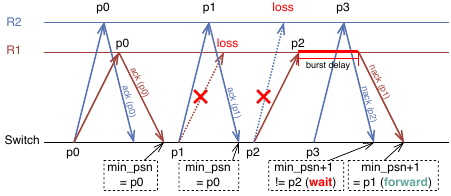}
	\caption{An example of NACK filtering.}
	\label{fig:nack-delay}
	\vspace{-0.5cm}
\end{figure}

We illustrate an example in Fig.~\ref{fig:nack-delay}. There are two receivers and one switch. The switch generates two copies of data and forwards them to two receivers. There are two lost packet, $p1_{R1}$ and $p2_{R2}$, and two associated NACK packets, $nack_{p1}$ and $nack_{p2}$. The switch should forward the $nack_{p1}$ because it contains the minimum expected PSN. If the $nack_{p2}$ is forwarded to the sender first, the loss of $p1_{R1}$ will be covered because the sender will assume that all the packets before the expected PSN of $nack_{p2}$ (\ie, $p2$) have been received. Thus the sender won't retransmit $p1$ anymore, and the reliability is compromised.

Therefore, the NACK packet should be forwarded only when all receivers have acknowledged all packets with PSN smaller than its expected PSN. We implement this judgement in Line~\ref{line:nack} of Algorithm~\ref{alg:generation}. If the condition is not satisfied, the switch keeps waiting, during which new ACK/NACK packets keep coming. If the switch receives a new NACK and its PSN is not great than the recorded $T.nack.ePSN$, the $T.nack.ePSN$ is updated, and the NACK generation condition is rechecked, as shown in Line~\ref{line:nackupdate} of Algorithm~\ref{alg:ack-aggre}.

\subsection{Other Considerations} \label{other-design}

\parab{From single source to multiple sources.}
Multicast source switching is common in datacenter communications. For example, in the \textit{PB} phase of HPL, different nodes play as the multicast source in different epochs. \sys can support this communication requirement by switching multicast source inside the group without reestablishing QPs. In particular, when the last multicast source finishes transmission, the multicast members use the already established QPs for the subsequent multicast communication. The multicast forwarding table maintained by the switches stays almost untouched. In Appendix~\ref{apx:source-switch}, we provide a detailed description of multicast source switching.

\parab{Congestion control.} 
As described above, we reuse the entire RC transport at the end-host for RNIC compatibility. This means that we also reuse the built-in congestion control (CC) mechanism to regulate the multicast sending rate. Rationally, a multicast congestion control algorithm should match its sending rate with the most congested path (\ie, bottleneck) of its data distribution tree~\cite{widmer2001extending, rizzo2000pgmcc}.

To this end, we enhance the switch feedback aggregation with a congestion signal filtering mechanism while remaining the end-host CC unchanged. Specifically, we maintain a congestion counter for each link at the switch, recording the congestion signal from the receivers or downstream switches. We then perform signal filtering to only pass through the congestion signal from the most congested link\footnote{There are different implementations of congestion signals. Some adopt stand-alone CNP, while others reuse the ACK to carry the congestion bits. Our congestion signal filtering design works for them all.} (forming a path when cascading each switch link end-to-end). Further, a periodic aging mechanism is added to update the congestion counter to match the frequently changing network dynamics.

\section{Implementation} \label{imple}

\begin{figure*}[t]
	\centering
	\includegraphics[width=0.95\linewidth]{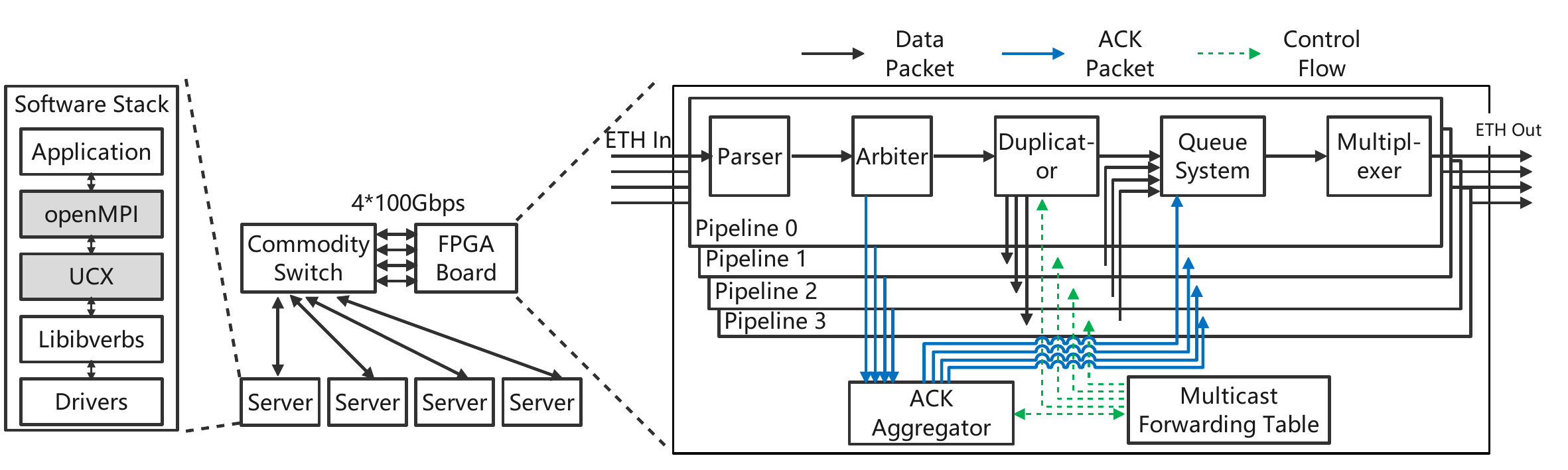}
	\caption{\sys testbed consists of a commodity switch, an FPGA-based prototype, and four servers. The FPGA board implements four processing pipelines, each capable of line-rate traffic processing for a 100Gbps Ethernet interface.}
	\label{fig:fpgaprototype}
	\vspace{-0.25cm}
\end{figure*}

\sys's implementation consists of (\romannumerber{1}) a fully-functional \sys switch prototype which implements the overall in-fabric logic; and (\romannumerber{2}) a set of software APIs exposed to applications. Our prototype is built upon a FPGA-assisted commodity switch.


\parab{FPGA-based prototype.} We implement the group registration, data packet duplication, header modification, and feedback aggregation logics on an FPGA board. The board is equipped with a commodity FPGA chip~\cite{ultrascale} and four 100Gbps Ethernet interfaces. The FPGA resource utilization is shown in Table~\ref{tab:overhead}. We build our testbed with the FPGA board, a commodity Ethernet switch, and four servers, as illustrated in Fig.~\ref{fig:fpgaprototype}. Each server is equipped with a commodity RNIC. The FPGA board and four RNICs are connected to the commodity switch through 100Gbps Ethernet interfaces. 

The commodity switch is configured by Access Control List (ACL) to route the servers' multicast traffic to the FPGA board. The FPGA board identifies the multicast data (ACK\footnote{In Fig.~\ref{fig:fpgaprototype}, we use ACK to represent all types of feedback.}) packets through the specific packet header by \textit{Parser} and \textit{Arbiter}. The data (ACK) packets will be duplicated (aggregated) by \textit{Duplicator} (\textit{ACK Aggregator}). The resulting packets will be pushed in \textit{Queue System}, waiting for the \textit{Multiplexer} to schedule in case for queue competition. Finally, the duplicated data (aggregated ACK) packets are sent back to the commodity switch. During processing, the \textit{Multicast Forwarding Table} is accessed when needed. 

\parab{Software APIs.} We provide various communication libraries and middleboxes for \sys multicast support. Take the commonly-used OpenMPI as an example, we modify the OpenMPI (v4.1.1)~\cite{openmpi} and UCX (v2.3)~\cite{ucx} to adapt to \sys's design, as shown in Fig.~\ref{fig:fpgaprototype}. Specifically, we add a new implementation of $MPI\_Bcast$ and modify UCX for multicast QPs creation and data transmission. When the new $MPI\_Bcast$ is called, the MPI process calls the UCX to establish QPs for multicast. Multicast members exchange their QPs information, and the handshake starts, as described in Appendix \ref{apx:regis}. Once the multicast group is successfully established, the UCX finally calls the RDMA primitives defined in the well-known \textit{libibverbs}~\cite{libibverbs} to transmit data. The software modifications at the end-host are transparent to the upper-layer applications and don't require any RNIC or RDMA driver modification.

\begin{table}[t]
	\small
    \centering
    \begin{tabular}{|p{0.2\linewidth}|p{0.18\linewidth}|p{0.18\linewidth}|p{0.18\linewidth}|}
    \hline
    \textbf{Resource} & \hfil \textbf{LUT} & \hfil \textbf{Register} & \hfil \textbf{BRAM} \\
    \hline
   	\textbf{Usage} & \hfil 53169 & \hfil 15391 & \hfil 188 \\
    \hline
    \end{tabular}
    \caption{Resource usage of the \sys in-fabric logic.}
    \label{tab:overhead}
    \vspace{-0.25cm}
\end{table}

\section{Evaluation}\label{eva}

We evaluate \sys's performance through extensive testbed and simulation experiments. In testbed experiments, we first examine \sys using the micro-benchmark ($\S$\ref{micro}); then we deploy two realistic applications ($\S$\ref{realapp}) upon \sys and show their performance improvement. The testbed's topology and configuration are described in $\S$\ref{imple}. In simulations ($\S$\ref{simu}), we evaluate \sys in an extended large-scale topology and explicitly estimate the goodput of \sys over different packet loss rate. Our experiment results reveal that:

\begin{itemize}
\vspace{-0.2cm}
\item \sys accelerates \mpibcast by up to 2.2$\times$ compared with OpenMPI.\vspace{-0.2cm}
\item \sys improves the performances of the realistic applications. Specifically, \sys speeds up the HPL communication by up to 2.9$\times$, improves the IOPS throughput of storage data replication by 2.7$\times$, and reduces the single IO latency by up to 65$\%$.\vspace{-0.2cm}
\item \sys achieves consistently high performance in large-scale topologies. Its communication speedup scales well with the multicast group size. Its performance gain is also resilient to packet losses. \vspace{-0.2cm}
\end{itemize}

\subsection{Micro-benchmark} \label{micro}
We first evaluate \mpibcast, one of the basic MPI primitives, through the modified OpenMPI~\cite{openmpi} and UCX~\cite{ucx}. We measure the job completion time (JCT) as the main metric. In our testbed, we select one server as \mpibcast source and three servers as \mpibcast receivers, forming a small one-to-three multicast group. The original OpenMPI performance is then compared with that of \sys. 

\begin{figure}[t]
	\centering
	\includegraphics[width=\linewidth]{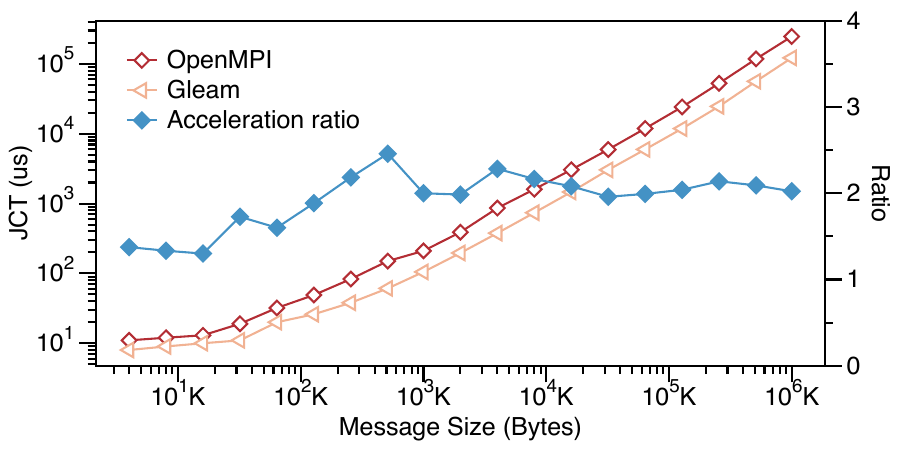}
	\caption{The JCTs of \mpibcast. The left vertical axis indicates the multicast JCTs. The right vertical axis shows the acceleration ratios of \sys over OpenMPI.}
	\label{fig:exp:bcast}
	\vspace{-0.5cm}
\end{figure}

We measure the JCT under various multicast message sizes, and calculate \sys's acceleration radio, shown in Fig.~\ref{fig:exp:bcast}. \sys achieves lower JCTs across various message sizes compared with OpenMPI. With larger message size, \sys achieves lower JCT. For instance, \sys achieves 12$\mu s$ reduction (1.6$\times$ acceleration radio) with 64KB message and 124$ms$ reducetion (2$\times$ acceleration radio) with 1GB message. For message size larger than  128KB, \sys stably achieve about 50$\%$ less JCT. 

The performance improvment of \sys stems from its optimal bandwidth utilization. \sys delivers data efficiently without bandwidth wastage. In contrast, OpenMPI inevitably wastes bandwidth because it transmits identical data multiple times. Consequently, with larger message size, \sys saves more bandwidth and thus offers lower JCT.
Through these micro-benchmark experiments, we validate the correctness of our \sys prototype and demonstrate that \sys can speed up multicast communication under simple settings.

\subsection{Realistic Applications} \label{realapp}

We deploy two realistic applications with multicast patterns (\S\ref{multicast-pattern}): HPL (\S\ref{eval:hpl}) and storage data replication (\S\ref{eval:storage}); and evaluate their performances with and without \sys. 

\subsubsection{High-performance Linpack (HPL)} \label{eval:hpl}
We integrate \sys into HPL~\cite{hpl}, and build a HPL cluster using our prototype testbed to evaluate the performance of HPL. Each HPL epoch contains three steps:  \textit{Panel Factorization}, \textit{Panel Broadcast (PB)} and \textit{Update}. The \textit{PB} is a standard multicast transmission, and the \textit{Update} includes a communication phase, called \textit{Row Swap (RS)}, which can be implemented with multicast. The volume of multicast traffic is around several GBytes at the first epoch and linearly decrease to nearly zero as the computing proceeds.

We test HPL's \textit{PB} and \textit{RS} stages separately. We measure the total JCT (including computation and communication) of HPL when speeding up \textit{PB} and \textit{RS} solely, and only the communication time of \textit{PB} and \textit{RS}. Moreover, the \textit{RS} communication volume depends on the real-time computing result; thus we evaluate \textit{RS} with two data distributions, which are uniform and centralized. The original HPL implementation is selected as alternative solution compared with \sys, where the \textit{PB} and \textit{RS} are recommended to use the \textit{increasing-ring} (abbreviated as \textit{ring} in following HPL evaluations) and \textit{long} algorithms, respectively~\cite{hpl}. Both algorithms are overlay multicast solutions implemented by multiple underlay RC unicasts.

\begin{figure*}[t]
	\minipage[t]{0.3\textwidth}
  		\includegraphics[width=\linewidth]{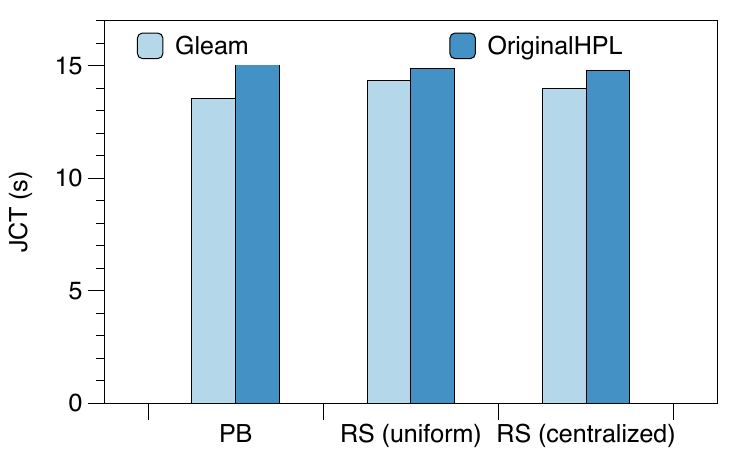}
  		\caption{HPL JCTs (computation $\&$ communication).}\label{fig:exp:hpl-total}
	\endminipage\hfill
	\minipage[t]{0.3\textwidth}
  		\includegraphics[width=\linewidth]{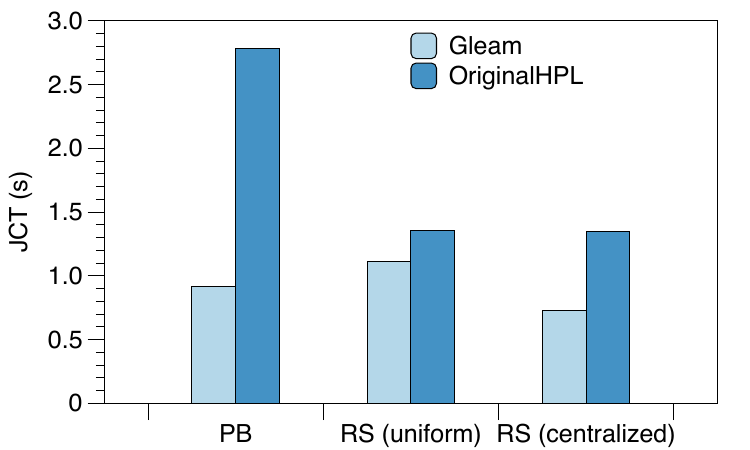}
  		\caption{HPL JCTs (communication only).}\label{fig:exp:hpl-commu}
	\endminipage\hfill
	\minipage[t]{0.3\textwidth}
  		\includegraphics[width=\linewidth]{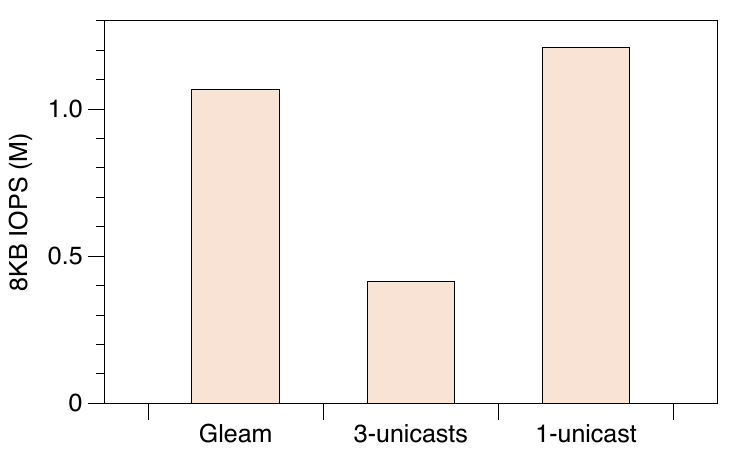}
  		\caption{Writing throughput in IOPS.}\label{fig:exp:write-thro}
	\endminipage\hfill
\end{figure*}

As results illustrated in Fig.~\ref{fig:exp:hpl-total} and Fig.~\ref{fig:exp:hpl-commu}, \sys reduces JCT in all settings. Specifically, with communication time only, \sys reduces the JCT of \textit{PB}, \textit{RS (uniform)}, and \textit{RS (centralized)} by 67$\%$, 18$\%$, and 46$\%$, respectively. With both computation and communication time included, \sys reduces the JCT of \textit{PB}, \textit{RS (uniform)}, and \textit{RS (centralized)} by up to 12$\%$, 4.67$\%$, and 9.55$\%$, respectively.

\parab{Tolerance of Data Distribution} Note that \sys achieves better improvement in non-random data distribution (46$\%$) compared with uniform data distribution (18$\%$). This is expected as the performance of \sys doesn't depend on data distribution. However, the \textit{long} algorithm is sensitive to data distribution. The \textit{long} algorithm achieves best performance when data is uniformly distributed, but degrades if the data is non-random distributed or even centralized, as more communication is needed to make the data evenly distributed before data exchanging between neighbors. 

\subsubsection{Storage Data Replication}\label{eval:storage}
We also integrate \sys into a proprietary commodity distributed storage system. We utilize our testbed prototype to evaluate the performance of storage data replication. We select one node as client and three nodes as servers, thus forming a 3-copies writing setting. We compare the performance of \sys with 3-unicasts, and the one-copy writing is also measured as a baseline. The \rdwrite operation is used in all solutions. In 3-unicasts, client maintains three RC connections with three servers. In \sys, client only maintains one RC connection. We measure the writing throughput using IOPS (IO per second) as the main metric, and the single IO latency.

\parab{Throughput.} We set IO size as 8KB and let the client keep writing data to three servers. We measure the average IOPS achieved by the client. As Fig.~\ref{fig:exp:write-thro} shows, \sys can achieve nearly optimal writing IOPS, about 1.167M IOPS, which is comparable with the ideal one-copy writing's 1.188M IOPS. On the contrary, the 3-unicasts only achieves 0.413M IOPS, only about 35$\%$ of \sys. With \sys, the application goodput can achieve 76.5Gbps (1.1M $\times$ 8KB), while 3-unicasts only can reach 26.24Gbps. This is because the client only need to send one copy of data for 3-replications writing with \sys, effectively mitigating the bandwidth bottleneck at the client's output link. 

\begin{figure}[t]
	\centering
	\includegraphics[width=0.8\linewidth]{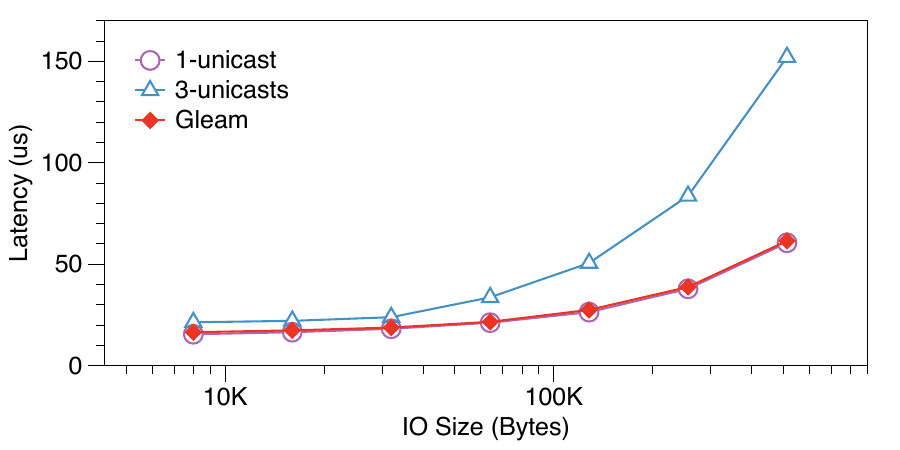}
	\caption{Single IO latency.}
	\label{fig:exp:write-latency}
	\vspace{-0.5cm}
\end{figure}

\parab{IO Latency.} In addition, we measure the single IO latency over different IO size, shown in Fig.~\ref{fig:exp:write-latency}. The single IO latency is defined as the period between the client submits the \rdwrite request and receives the CQE from the RNIC. The result shows that \sys achieves significantly lower latency than 3-unicasts, and delivers a comparable latency with the ideal one-copy. For instance, \sys reduces IO latency by about 40$\%$ and 60$\%$ in 64KB and 512KB, compared with 3-unicasts. Moreover, \sys accomplishes lower latency in larger IO size. As result shown, the gap between \sys and 3-unicasts is enlarged as IO size increasing.

The advantage of \sys comes from the reduction of end-host storage stacks involvement and the total transmitted traffic volume. 3-unicasts transmits the identical data and experiences the same storage stacks three times. As a result, the IO latency is increased. On the other hand, \sys enables the client to run the storage stack and transmit the data only once, which effectively reduces the IO latency.

\subsection{Simulations} \label{simu}
We provide complementary experiments using ns-3~\cite{ns3}. We first build a large-scale topology and evaluate \sys over different multicast scales. We then simulate a lossy environment and measure \sys over different packet loss rates.

\parab{Simulation setting.}
We simulate a large-scale 3-layer fat-tree topology with 16384 servers and a 1:1 oversubscription ratio. Each server is equipped with one 200Gbps RNIC, and each switch has 64 200Gbps ports. 

\parab{Workload and Metrics.} We again use the HPL pattern as the simulation workload. We provide various workload scales, each labeled as $N*N$, meaning that $N*N$ nodes form a logical $N*N$ matrix to run HPL. In each workload, each row node would perform one \textit{PB}, and each column node would perform one \textit{RS}. With \sys, the \textit{PB} procedure involves $N$ multicast groups transmitting simultaneously, each consisting of $N$ group members. The \textit{RS} involves another different $N$ multicast groups. The \textit{ring} algorithm for \text{PB} and \textit{long} algorithm for \textit{RS} are simulated as comparisons with \sys. We use the total JCT, \ie, the total communication time of \textit{PB} and \textit{RS}, as the experiment metric.  

The simulations are time-costly because of its large topology and the more than four million flows. To accelerate the simulation, we integrate MPI framework into the ns-3 simulator to enable the parallel simulation with multiple threads. The simulation run time is reduced by 70\% with 12 threads.

\begin{figure*}[t]
	\minipage[t]{0.32\textwidth}
  		\includegraphics[width=\linewidth]{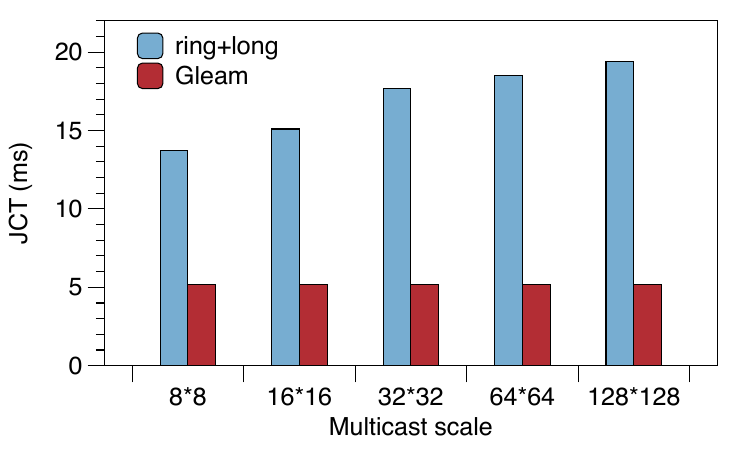}
  		\caption{JCTs with different multicast scales.}\label{fig:exp:large-scale}
	\endminipage\hfill
	\minipage[t]{0.32\textwidth}%
  		\includegraphics[width=\linewidth]{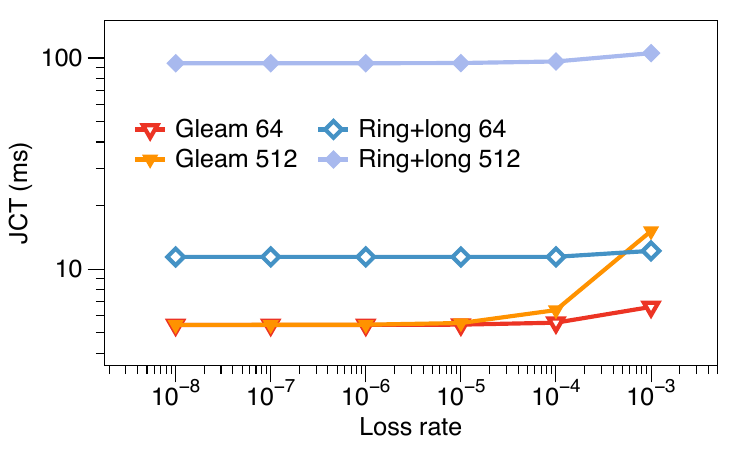}
  		\caption{JCTs under various packet loss rates.}\label{fig:exp:tct-loss}
  	\endminipage\hfill
  	\minipage[t]{0.32\textwidth}
  		\includegraphics[width=\linewidth]{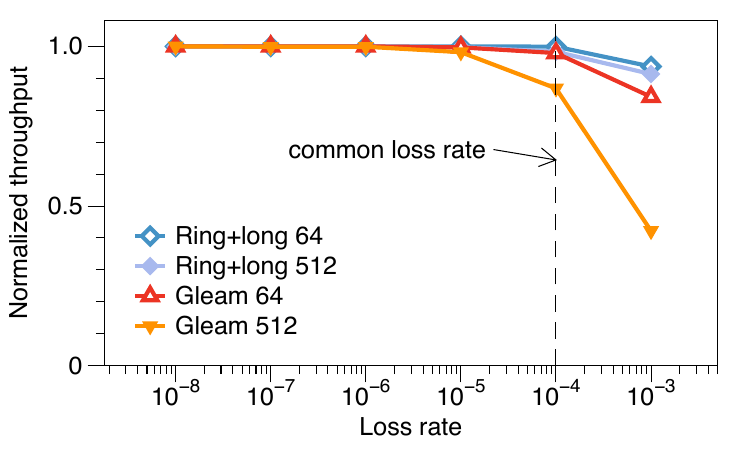}
  		\caption{Normalized throughput (goodput) under various packet loss rates.}\label{fig:exp:thro-loss}
	\endminipage
	\vspace{-0.5cm}
\end{figure*}

\parab{Large-scale multicast.}
We measure the average JCT of HPL workload over different multicast scales. In particular, we select five scales: $8*8$, $16*16$, $32*32$, $64*64$, $128*128$. For each scale, we run multiple times and calculate the average JCT. The original HPL implementation (\textit{ring} for \text{PB} and \textit{long} for \textit{RS}) is compared with \sys. Results in Fig.~\ref{fig:exp:large-scale} show that \sys achieves lower JCT in all multicast scales with a reduction from 62$\%$ to 73$\%$. For instance, \sys{} reduces the JCT from 13.7ms to 5.17ms (62$\%$ reduction) over $8*8$ scale, and from 19.4ms to 5.17ms (73$\%$ reduction) over $128*128$ scale. The improvement increases with the growing multicast scale, consistent with the results in $\S$\ref{micro}.

Note that \sys's JCT doesn't scale up with increasing multicast scale. The reason is that, in \sys, the larger multicast scale only brings more parallel-transmitted multicast groups, while the traffic volume transmitted by each group stays the same. Because of \sys's group-level load-balancing (Appendix~\ref{apx:regis}), more multicast groups don't incur much congestion deterioration. In contrast, with the \textit{ring} and \textit{long} algorithms, the number of parallel unicast flows expands linearly with the growing multicast scale, which causes congestion and results in JCT degradation.

\parab{Loss tolerance.}
We compare the JCT of HPL workload under different packet loss rates, from $10^{-8}$ to $10^{-3}$, which are emulated via randomly discarding packets in the middle switches. We choose two group sizes (64 and 512) to evaluate. As shown in Fig.~\ref{fig:exp:tct-loss}, \sys{} presents good loss tolerance as it can maintain lower JCT compared with original HPL implementation (\textit{ring} and \textit{long}) under all loss rates. In particular, \sys{} reduces the JCT by 46\%-52\% with size 64 and 86\%-94\% with size 512, respectively. 

To better show \sys's reaction to loss, we calculate the goodput under different packet loss rates, \ie, the normalized throughput compared with setting without loss, shown in Fig.~\ref{fig:exp:thro-loss}. \sys's goodput decreases largely than \textit{ring} and \textit{long} because the multicast sender is responsible for multiple receivers, \ie, when any receiver loses a packet, the sender must retransmit it. In contrast, the sender in \textit{ring} and \textit{long} is only responsible for one receiver. 

Even though \sys is more sensitive to packet loss, \sys imposes at most 10\% goodput degradation with a loss rate not larger than 0.01\%, which is a common packet loss rate in data center~\cite{zhu2015congestion}. Even under an excessive 0.1\% loss rate, \sys can maintain 42\% goodput and perform better (gains a 7$\times$ lower JCT) than the original HPL implementation.

\section{Related Works} \label{works}
\parab{Internet and Datacenter multicast.} Multicast has been widely applied in large-scale Internet applications, such as Internet broadcast \cite{iptv}, video conferencing \cite{chen2011celerity}, and multiplayer games \cite{cho2009enabling}, \etc Prior works for the Internet \cite{chiang2018online, huang2016multicast,diab2020oktopus,ren2018optimal, diab2022yeti} mostly focus on the multicast routing, \ie, to find promising multicast paths, inside ISPs. For instance, Yeti~\cite{diab2022yeti} supports multicast routing with traffic engineering and service chaining requirements for large-scale ISPs. Yeti creates labels representing forwarding information for multicast graphs and processes these labels to forward packets to targeted paths. Although there are a bunch of prior works on the Internet, most of them merely provide best-effort delivery, which only works for applications without reliability requirement. 

There are some works~\cite{widmer2001extending, rizzo2000pgmcc} aim to provide reliability for datacenter applications upon approaches with best-effort delivery. However, existing reliable multicast solutions mainly adopt a TCP-like software stack and cannot meet the demand for high-speed communication in datacenters. In contrast, \sys  leverages the advanced RDMA stacks to process multicast traffic, providing high-speed reliable communication.


\parab{Multicast scalability.}
Datacenter applications impose a demand for high scalability. As the traditional IP multicast~\cite{crowcroft1988multicast}, along with its native group management, IGMP and tree construction protocol, PIM~\cite{estrin1998protocol}, are poor in scalability, many works~\cite{shahbaz2019elmo, diab2022orca, li2013scaling} attempt to address the scalability issue, \ie, supporting as much as possible multicast groups. For example, Elmo~\cite{shahbaz2019elmo} encodes the routing link of a multicast tree into rules formatted as packet header. Thus Elmo switch only needs to maintain rule parsing logic, reducing the total switch-maintained states. Orca~\cite{diab2022orca} utilizes the large memory space of the server, making servers assist in forwarding packets, reducing the switch's burden on maintaining states. 

These works that address the scalability issue are orthogonal with the \sys design. Our goal in this work is to provide a general multicast protocol with prominent RDMA features and reliability guarantee rather than compressing the switch-maintained states. As mentioned before, \sys can support at least 1K multicast groups using 0.92MB space, which is acceptable for a majority of multicast applications in datacenters. \sys can support even more multicast groups when getting extended further upon these works.


\section{Conclusion}
We present \sys, an RDMA-accelerated multicast protocol that significantly facilitates multicast communication while maintaining compatibility with commodity RNICs. \sys reuses existing RDMA RC transport to process multicast traffic, thus preserving the optimal bandwidth utilization and benefiting from superior RDMA functionality. \sys replaces packet headers and aggregates feedback in the fabric to remain compatible between existing RC logic and multicast traffic. We provide a fully functional \sys prototype, which requires no RNIC modification. Extensive testbed experiments and simulations demonstrate \sys's superior performance in multicast acceleration.

\sys opens the door for efficiently leveraging the prominent RDMA stacks with in-fabric assistance to accelerate group communication patterns. While this work mainly focuses on multicast; for future works, we plan to extend \sys for more group communication patterns, such as many-to-one (\eg, $MPI\_Reduce$) and many-to-many (\eg, $MPI\_Alltoall$), \etc


\bibliographystyle{plain}
\bibliography{main.bib}

\begin{thebibliography}{10}

\bibitem{Infiniband}
Inﬁniband architecture volume 1, general speciﬁcations, release 1.2.1.
\newblock \url{https://www.infinibandta.org/specs}, 2008.

\bibitem{rocev2}
Supplement to inﬁniband architecture speciﬁcation volume 1 release 1.2.2
  annex a17: Rocev2 (ip routable roce).
\newblock \url{https://www.infinibandta.org/specs}, 2014.

\bibitem{hpl}
Hpl - a portable implementation of the high-performance linpack benchmark for
  distributed-memory computers.
\newblock \url{https://netlib.org/benchmark/hpl/}, 2018.

\bibitem{mlsl}
Intel machine learning scalability library (mlsl).
\newblock \url{https://github.com/intel/MLSL}, 2019.

\bibitem{libibverbs}
libibverbs.
\newblock
  \url{https://github.com/linux-rdma/rdma-core/blob/master/Documentation/libibverbs.md},
  2021.

\bibitem{qpmodi}
Linux manual page.
\newblock \url{https://man7.org/linux/man-pages/man3/ibv_modify_qp.3.html/},
  2021.

\bibitem{iptv}
Bt iptv.
\newblock \url{https://bit.ly/3ssCvTz.}, 2022.

\bibitem{cx5}
Connectx-5.
\newblock \url{https://www.nvidia.com/en-us/networking/ethernet/connectx-5/},
  2022.

\bibitem{ns3}
ns-3, a discrete-event network simulator for internet systems.
\newblock \url{https://www.nsnam.org/}, 2022.

\bibitem{nccl}
Nvidia collective communication library (nccl).
\newblock \url{https://developer.nvidia.com/nccl}, 2022.

\bibitem{oneccl}
oneapi collective communications library (oneccl).
\newblock \url{https://github.com/oneapi-src/oneCCL}, 2022.

\bibitem{openmpi}
Openmpi: Open source high performance computing.
\newblock \url{https://www.open-mpi.org/}, 2022.

\bibitem{top500}
Top 500 list.
\newblock \url{https://www.top500.org/lists/top500/2022/06/}, 2022.

\bibitem{ucx}
Unifiex communication x.
\newblock \url{https://openucx.org/}, 2022.

\bibitem{ultrascale}
Virtex ultrascale.
\newblock
  \url{https://www.xilinx.com/products/silicon-devices/fpga/virtex-ultrascale.html/},
  2022.

\bibitem{chen2011celerity}
Xiangwen Chen, Minghua Chen, Baochun Li, Yao Zhao, Yunnan Wu, and Jin Li.
\newblock Celerity: A low-delay multi-party conferencing solution.
\newblock In {\em Proceedings of the 19th ACM international conference on
  Multimedia}, pages 493--502, 2011.

\bibitem{chiang2018online}
Sheng-Hao Chiang, Jian-Jhih Kuo, Shan-Hsiang Shen, De-Nian Yang, and Wen-Tsuen
  Chen.
\newblock Online multicast traffic engineering for software-defined networks.
\newblock In {\em IEEE INFOCOM 2018-IEEE Conference on Computer
  Communications}, pages 414--422. IEEE, 2018.

\bibitem{cho2009enabling}
Tae~Won Cho, Michael Rabinovich, KK~Ramakrishnan, Divesh Srivastava, and Yin
  Zhang.
\newblock Enabling content dissemination using efficient and scalable
  multicast.
\newblock In {\em IEEE INFOCOM 2009}, pages 1980--1988. IEEE, 2009.

\bibitem{mpiusage}
Sudheer Chunduri, Scott Parker, Pavan Balaji, Kevin Harms, and Kalyan Kumaran.
\newblock Characterization of mpi usage on a production supercomputer.
\newblock In {\em SC18: International Conference for High Performance
  Computing, Networking, Storage and Analysis}, pages 386--400. IEEE, 2018.

\bibitem{crowcroft1988multicast}
Jon Crowcroft and Karen Paliwoda.
\newblock A multicast transport protocol.
\newblock In {\em Symposium proceedings on Communications architectures and
  protocols}, pages 247--256, 1988.

\bibitem{dct}
D~Crupnico, M~Kagan, A~Shahar, N~Bloch, and H~Chapman.
\newblock Dynamically-connected transport service, may 19 2011.
\newblock {\em URL https://www. google. com/patents/US20110116512. US Patent
  App}, 12(621,523).

\bibitem{dean2004mapreduce}
Jeffrey Dean and Sanjay Ghemawat.
\newblock Mapreduce: Simplified data processing on large clusters.
\newblock 2004.

\bibitem{diab2022yeti}
Khaled Diab and Mohamed Hefeeda.
\newblock Yeti: Stateless and generalized multicast forwarding.
\newblock In {\em 19th USENIX Symposium on Networked Systems Design and
  Implementation (NSDI 22)}, pages 1093--1114, 2022.

\bibitem{diab2020oktopus}
Khaled Diab, Carlos Lee, and Mohamed Hefeeda.
\newblock Oktopus: Service chaining for multicast traffic.
\newblock In {\em 2020 IEEE 28th International Conference on Network Protocols
  (ICNP)}, pages 1--11. IEEE, 2020.

\bibitem{diab2022orca}
Khaled Diab, Parham Yassini, and Mohamed Hefeeda.
\newblock Orca: Server-assisted multicast for datacenter networks.
\newblock In {\em 19th USENIX Symposium on Networked Systems Design and
  Implementation (NSDI 22)}, pages 1075--1091, 2022.

\bibitem{dongarra2003linpack}
Jack~J Dongarra, Piotr Luszczek, and Antoine Petitet.
\newblock The linpack benchmark: past, present and future.
\newblock {\em Concurrency and Computation: practice and experience},
  15(9):803--820, 2003.

\bibitem{dragojevic2014farm}
Aleksandar Dragojevi{\'c}, Dushyanth Narayanan, Miguel Castro, and Orion
  Hodson.
\newblock $\{$FaRM$\}$: Fast remote memory.
\newblock In {\em 11th USENIX Symposium on Networked Systems Design and
  Implementation (NSDI 14)}, pages 401--414, 2014.

\bibitem{estrin1998protocol}
Deborah Estrin, Dino Farinacci, Ahmed Helmy, David Thaler, Stephen Deering,
  Mark Handley, Van Jacobson, Ching-Gung Liu, Puneet Sharma, and Liming Wei.
\newblock Protocol independent multicast-sparse mode (pim-sm): Protocol
  specification.
\newblock Technical report, 1998.

\bibitem{fenner1997internet}
William Fenner.
\newblock Internet group management protocol, version 2.
\newblock Technical report, 1997.

\bibitem{gao2021cloud}
Yixiao Gao, Qiang Li, Lingbo Tang, Yongqing Xi, Pengcheng Zhang, Wenwen Peng,
  Bo~Li, Yaohui Wu, Shaozong Liu, Lei Yan, et~al.
\newblock When cloud storage meets $\{$RDMA$\}$.
\newblock In {\em 18th USENIX Symposium on Networked Systems Design and
  Implementation (NSDI 21)}, pages 519--533, 2021.

\bibitem{gibiansky2017bringing}
Andrew Gibiansky.
\newblock Bringing hpc techniques to deep learning.
\newblock {\em Baidu Research, Tech. Rep.}, 2017.

\bibitem{guo2016rdma}
Chuanxiong Guo, Haitao Wu, Zhong Deng, Gaurav Soni, Jianxi Ye, Jitu Padhye, and
  Marina Lipshteyn.
\newblock Rdma over commodity ethernet at scale.
\newblock In {\em Proceedings of the 2016 ACM SIGCOMM Conference}, pages
  202--215, 2016.

\bibitem{huang2016multicast}
Liang-Hao Huang, Hsiang-Chun Hsu, Shan-Hsiang Shen, De-Nian Yang, and Wen-Tsuen
  Chen.
\newblock Multicast traffic engineering for software-defined networks.
\newblock In {\em IEEE INFOCOM 2016-The 35th Annual IEEE International
  Conference on Computer Communications}, pages 1--9. IEEE, 2016.

\bibitem{jiang2020unified}
Yimin Jiang, Yibo Zhu, Chang Lan, Bairen Yi, Yong Cui, and Chuanxiong Guo.
\newblock A unified architecture for accelerating distributed $\{$DNN$\}$
  training in heterogeneous $\{$GPU/CPU$\}$ clusters.
\newblock In {\em 14th USENIX Symposium on Operating Systems Design and
  Implementation (OSDI 20)}, pages 463--479, 2020.

\bibitem{kalia2014using}
Anuj Kalia, Michael Kaminsky, and David~G Andersen.
\newblock Using rdma efficiently for key-value services.
\newblock In {\em Proceedings of the 2014 ACM Conference on SIGCOMM}, pages
  295--306, 2014.

\bibitem{li2014communication}
Mu~Li, David~G Andersen, Alexander~J Smola, and Kai Yu.
\newblock Communication efficient distributed machine learning with the
  parameter server.
\newblock {\em Advances in Neural Information Processing Systems}, 27, 2014.

\bibitem{li2013scaling}
Xiaozhou Li and Michael~J Freedman.
\newblock Scaling ip multicast on datacenter topologies.
\newblock In {\em Proceedings of the ninth ACM conference on Emerging
  networking experiments and technologies}, pages 61--72, 2013.

\bibitem{massie2004ganglia}
Matthew~L Massie, Brent~N Chun, and David~E Culler.
\newblock The ganglia distributed monitoring system: design, implementation,
  and experience.
\newblock {\em Parallel Computing}, 30(7):817--840, 2004.

\bibitem{miao2022luna}
Rui Miao, Lingjun Zhu, Shu Ma, Kun Qian, Shujun Zhuang, Bo~Li, Shuguang Cheng,
  Jiaqi Gao, Yan Zhuang, Pengcheng Zhang, et~al.
\newblock From luna to solar: the evolutions of the compute-to-storage networks
  in alibaba cloud.
\newblock In {\em Proceedings of the ACM SIGCOMM 2022 Conference}, pages
  753--766, 2022.

\bibitem{mittal2015timely}
Radhika Mittal, Vinh~The Lam, Nandita Dukkipati, Emily Blem, Hassan Wassel,
  Monia Ghobadi, Amin Vahdat, Yaogong Wang, David Wetherall, and David Zats.
\newblock Timely: Rtt-based congestion control for the datacenter.
\newblock In {\em Proceedings of the 2015 ACM SIGCOMM Conference}, page
  537–550, 2015.

\bibitem{monga2021birds}
Sumit~Kumar Monga, Sanidhya Kashyap, and Changwoo Min.
\newblock Birds of a feather flock together: Scaling rdma rpcs with flock.
\newblock In {\em Proceedings of the ACM SIGOPS 28th Symposium on Operating
  Systems Principles}, pages 212--227, 2021.

\bibitem{ren2018optimal}
Bangbang Ren, Deke Guo, Guoming Tang, Xu~Lin, and Yudong Qin.
\newblock Optimal service function tree embedding for nfv enabled multicast.
\newblock In {\em 2018 IEEE 38th international conference on distributed
  computing systems (ICDCS)}, pages 132--142. IEEE, 2018.

\bibitem{rizzo2000pgmcc}
Luigi Rizzo.
\newblock pgmcc: a tcp-friendly single-rate multicast congestion control
  scheme.
\newblock {\em ACM SIGCOMM Computer Communication Review}, 30(4):17--28, 2000.

\bibitem{shahbaz2019elmo}
Muhammad Shahbaz, Lalith Suresh, Jennifer Rexford, Nick Feamster, Ori
  Rottenstreich, and Mukesh Hira.
\newblock Elmo: Source routed multicast for public clouds.
\newblock In {\em Proceedings of the ACM Special Interest Group on Data
  Communication}, pages 458--471. 2019.

\bibitem{shi2016fast}
Jiaxin Shi, Youyang Yao, Rong Chen, Haibo Chen, and Feifei Li.
\newblock Fast and concurrent $\{$RDF$\}$ queries with $\{$RDMA-Based$\}$
  distributed graph exploration.
\newblock In {\em 12th USENIX Symposium on Operating Systems Design and
  Implementation (OSDI 16)}, pages 317--332, 2016.

\bibitem{widmer2001extending}
J{\"o}rg Widmer and Mark Handley.
\newblock Extending equation-based congestion control to multicast
  applications.
\newblock In {\em Proceedings of the 2001 conference on Applications,
  technologies, architectures, and protocols for computer communications},
  pages 275--285, 2001.

\bibitem{zhu2015congestion}
Yibo Zhu, Haggai Eran, Daniel Firestone, Chuanxiong Guo, Marina Lipshteyn,
  Yehonatan Liron, Jitendra Padhye, Shachar Raindel, Mohamad~Haj Yahia, and
  Ming Zhang.
\newblock Congestion control for large-scale rdma deployments.
\newblock {\em ACM SIGCOMM Computer Communication Review}, 45(4):523--536,
  2015.

\end{thebibliography}

\appendix

\section{More about Multicast Table Registration} \label{apx:regis}
The in-fabric logic of \sys is based on the switches' multicast forwarding table. \sys follows the table information to copy packets and forward them to specific output ports. Although existing multicast solutions commonly adopt the table-based approach, \eg, IP multicast~\cite{crowcroft1988multicast}, the multicast table registration in \sys has some primary differences from them.

The first difference is that the table registration of \sys is centralized, while the traditional works~\cite{fenner1997internet,crowcroft1988multicast, estrin1998protocol} perform a distributed registration algorithm\cite{rfc3376}. The critical insight pushing registration from distributed to centralized is that the datacenter is highly autonomous, and the multicast membership is highly controlled. This centralized approach is widely used in various datacenter frameworks~\cite{shahbaz2019elmo, diab2022orca}. Secondly, previous works only register layer-3 states to switch. \sys, on the other hand, registers both layer-3 states and layer-4 states to switches for the connectivity and reliability supports, as introduced in $\S$\ref{design}.

\begin{figure}[t]
	\centering
	\includegraphics[width=0.95\linewidth]{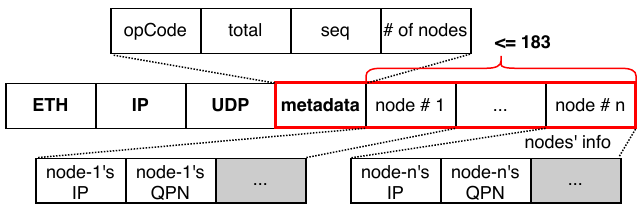}
	\caption{The \envelope packet format.}
	\label{fig:envelope-packet}
	\vspace{-0.25cm}
\end{figure}

\sys's registration protocol is centralized and relied on an application-assigned master node. Before communication starts, the master node in the multicast group collects the connection states (including the layer-3 IP and layer-4 IB information) of all other members through an out-of-band protocol (\eg, TCP). Then, the master node fits these states into the self-developed \envelope protocol's packet layout, which is illustrated in Fig.~\ref{fig:envelope-packet}. The \envelope packet is identified by the destination IP (\ie, GroupIP) and a specific UDP port number. The payload contains metadata and detailed connection states of each node, including the node's IP and QPN. The metadata comprises group statistics, where $seq$ and $total$ indicate the sequence and the total number of \envelope packets. Limited by the MTU (typically 1500Bytes), one \envelope packet can contain at most 183 nodes; thus, the connection states of a multicast group with more than 183 members must span multiple \envelope packets.

\begin{figure}[t]
	\centering
	\includegraphics[width=0.8\linewidth]{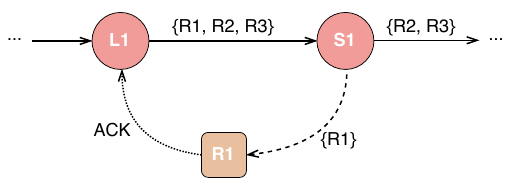}
	\caption{An example of \envelope packets transmission.}
	\label{fig:envelope-transmission}
	\vspace{-0.5cm}
\end{figure}

Upon receiving the \envelope packet, the switch builds its local multicast forwarding table and sends one or more new \envelope packets to downstream devices. Algorithm~\ref{alg:three} illustrates the behavior of \sys switches. An \envelope packet ($p$) carries a GroupIP and an array of multicast member connection states ($p.array$). The switch first generates an empty table indexed by $p.groupIP$. Then the switch iterates over $p.array$ to append items to the created table. For every node in the array, the switch finds this node's routing information through the normal unicast routing table. If this node is directly connected (\eg, connected to $port_i$), then the switch creates a table item with $port$ as $i$, marks this item's type to $connected$, and fills this node's connection states into this item. Otherwise, if this node isn't directly connected, the switch finds the set of possible output ports ($set_p$) for this node. If one port in $set_p$ has been marked as $forwarded$, the switch selects this port again to distribute data optimally and save bandwidth. If all ports in $set_p$ are new, the switch selects the least utilized port among $set_p$ to perform a group-level load balancing. 

After processing \envelope packet and filling the multicast forwarding table, the switch generates one or more new \envelope packets to ports included in the table. The new \envelope packet that through each port only contains connection states of nodes that select this port. We show the actions that taken by $S_1$ in Fig.~\ref{fig:envelope-transmission}, and the whole topology is shown in Fig.~\ref{fig:overview}. The \envelope packet received by $S_1$ contains $R_1$, $R_2$, and $R_3$. As instructed by Algorithm~\ref{alg:three}, $S_1$ should forward a data copy to $port_{L2}$ (eventually reaching $R_1$), and a copy to $port_{C2}$ (eventually reaching $R_2$ and $R_3$). Thus the \envelope packets forwarded to $port_{L2}$ and $port_{C2}$ contains information of $\{R_1\}$ and $\{R_2, R_3\}$, respectively, used in downstream devices to build their local forwarding table.
   
Finally, if a node receives an \envelope packet, and its IP address is included in the packet, this node will answer an ACK back to the master node to confirm its participation. After the master node collects all nodes' confirmation ACKs, the control-plane table registration is finished, and the multicast transmission can start. 

\begin{algorithm}[t]
\caption{Multicast Forwarding Table Registration}\label{alg:three}
\begin{algorithmic}[1]
\State $p\gets $ received \envelope packet
\State $n\gets$ the number of switch ports
\State create multicast forwarding table $T[p.groupIP]$
\State $R[n]\gets \{ 0 \}$ \Comment{\textcolor{gray}{for creating new \envelope packets}}
\For{$node$ in $p$}\Comment{\textcolor{gray}{loop over nodes in $p$ and build $T$}}
	\If{$node$ is directly connected to port $i$}
		\State $out \gets i$ \Comment{\textcolor{gray}{create new entry}}
		\State $Entry \gets $ a new entry
		\State $Entry.port \gets out$
		\State $Entry.type \gets connected$
		\State $Entry.vaue \gets $ node's connection states
		\State $T[p.groupIP].append(Entry)$
	\Else
		\State $S\gets $ the set of accessible ports
		\If{port $j\in S$ has been marked as $forwarded$}
			\State $out \gets j$ \Comment{\textcolor{gray}{reuse exiting entry}}
		\Else \Comment{\textcolor{gray}{create new entry}}
			\State $out \gets $ the least utilized port in $S$
			\State $Entry \gets $ a new entry
			\State $Entry.port \gets out$
			\State $Entry.type \gets forwarded$
			\State $T[p.groupIP].append(Entry)$
		\EndIf
	\EndIf
	\State update port utilization
	\State $R[out].append(node)$
\EndFor
\State \textcolor{gray}{// create new \envelope packets and send out}
\For{$Entry \in T[p.groupIP]$}
\State create \envelope packet $\overline{p}$ that contains nodes in $R[Entry.port]$
	\State send $\overline{p}$ through $Entry.port$
\EndFor
\end{algorithmic}
\end{algorithm}

\begin{figure}[t]
	\centering
	\includegraphics[width=\linewidth]{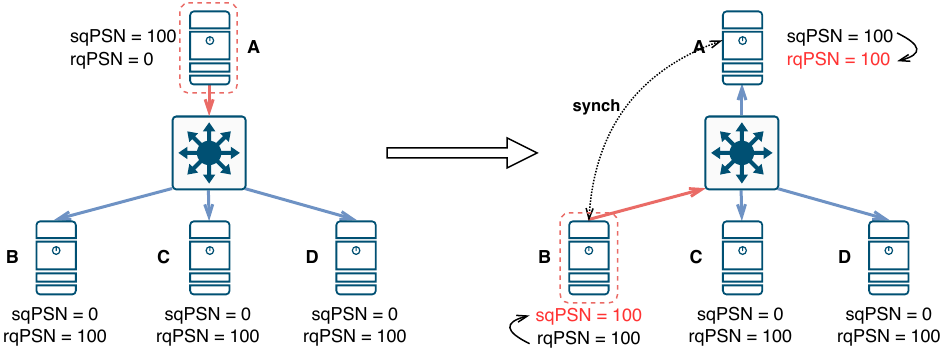}
	\caption{An example of multicast source switching}
	\label{fig:source-change}
	\vspace{-0.5cm}
\end{figure}

\section{Multicast Source Switching} \label{apx:source-switch}
\sys supports the source switching inside a multicast group. When the source of a multicast group changes, the switches can detect this by recognizing the change of the incoming port of multicast data packets. The new incoming port is recorded for later ACK forwarding. There are no other modifications to \sys's in-fabric logic. 

For the end-host, the old and new source nodes need to run a PSN synchronization procedure. Note that each QP maintains two PSN records. The \textit{Send Queue PSN} (sqPSN) is used to record the output packets, and the \textit{Receive Queue PSN} (rqPSN) is used to verify the input packets. The senders' sqPSN should equal the receiver's rqPSN at the beginning of the transmission. A PSN synchronization is needed to maintain this PSN consistency when the multicast source changes. For example, as shown in Fig.~\ref{fig:source-change}, node \textit{A} has multicasted 100 packets to nodes \textit{B}, \textit{C}, and \textit{D}. Assume that all nodes' sqPSN and rqPSN start from 0, the sqPSN of \textit{A} and rqPSNs of \textit{B}, \textit{C}, \textit{D} become 100 when the transmission ends. When the multicast source switches to \textit{B}, if \textit{B} starts transmission immediately, the PSN of sent packets would be \textit{B}'s current sqPSN, \ie, 0. These packets would be dropped by \textit{C} and \textit{D} as their rqPSNs are already 100. Therefore, the old and new multicast source nodes need to synchronize their PSNs. In particular, the old source node assigns its rqPSN as its sqPSN, and the new source node assigns its sqPSN as its rqPSN. This problem can also be avoided by using Dynamic Connected Transport (DCT)\cite{dct}, which synchronizes the PSN of the sender and the receiver with the DC Connect packet.

\section{Optimization for one-to-many \rdwrite} \label{apx:opwrite}
As mentioned in $\S$\ref{connection-handle}, to support one-to-many \rdwrite, we need an additional message to carry the MR info of different receivers for each \rdwrite request. This incurs extra bandwidth overhead, especially when the number of receivers is large. A straight idea to eliminate this overhead is to enable the possibility to write all receivers with the same MR info. We claim this idea is logically reasonable and easy to implement with few modifications to the RNICs. Note that the VA in MR is \textit{virtual} and is translated to the physical memory address by the RNIC. Therefore, it is reasonable to use an identical virtual address for all nodes, which can be translated to different physical addresses at different nodes. On the other hand, the \rkey is mainly used as an index for the RNIC to look up the Memory Translation Table (MTT). All receivers in a multicast group can share the same \rkey for the corresponding MR if some indexes are reserved for this purpose.

Currently, the VA is assigned by the OS kernel, and the \rkey is obtained from the RNIC. If the RNIC is modified so that the application can assign the VA and \rkey for an MR at the QP establishment phase, we can enable all receivers to use the same VA and \rkey. Consequently, the switch no longer needs to modify the MR info for different receivers, and the overhead for MR info update is avoided.


\end{document}